\documentclass[journal]{IEEEtran}

\usepackage{ifpdf}
\usepackage{url}
\ifpdf
\else
\fi

\usepackage{cite}
\usepackage{balance}
\usepackage{bm,comment,color}

\ifCLASSINFOpdf
  \usepackage[pdftex]{graphicx}
  \graphicspath{{../pdf/}{../jpeg/}}
  \DeclareGraphicsExtensions{.pdf,.jpeg,.png}
\else
  \usepackage[dvips]{graphicx}
  \graphicspath{{../eps/}}
  \DeclareGraphicsExtensions{.eps}
\fi
\usepackage{amsmath}

\usepackage{algorithm}
\usepackage{algpseudocode}
\algtext*{EndWhile}
\algtext*{EndIf}
\algtext*{EndFor}

\usepackage{array}
\usepackage{amsfonts} 
\usepackage{amssymb}
\usepackage{esint} 
\usepackage{units}
\usepackage{multirow}
\usepackage{comment}

\usepackage{color}
\newcommand{\red}[1]{{\color{red}{#1}}} 
\newcommand{\blue}[1]{{\color{blue}{#1}}} 

\newcommand{\yellow}[1]{{\color{yellow}{#1}}} 
 
\allowdisplaybreaks

\usepackage{pgfplots}
\usepackage{tikz}
\usetikzlibrary{calc}
\makeatletter
\newcommand{\gettikzxy}[3]{%
  \tikz@scan@one@point\pgfutil@firstofone#1\relax
  \edef#2{\the\pgf@x}%
  \edef#3{\the\pgf@y}%
}
\usetikzlibrary{spy,backgrounds}
\pgfplotsset{compat=newest}
\usetikzlibrary{plotmarks}
\usetikzlibrary{arrows.meta}
\usepgfplotslibrary{patchplots}
\usepackage{grffile}
\newlength\fheight 
\newlength\fwidth 
\usepgfplotslibrary{fillbetween}

\usepackage{acronym}

\usepackage{tabu,longtable}
\usepackage{subfigure}

\usepackage{stfloats}
\usepackage[hidelinks]{hyperref}
\usepackage{xcolor}
\acrodef{3gpp}[3GPP]{3rd Gneration Partnership Project}
\acrodef{6d}[6D]{six-dimensional}
\acrodef{ad}[AD]{autonomous drive}
\acrodef{adas}[ADAS]{advanced driver assistance system}
\acrodef{aoa}[AOA]{angles-of-arrival}
\acrodef{aod}[AOD]{angles-of-departure}
\acrodef{aosa}[AOSA]{array-of-subarray}
\acrodef{ap}[AP]{access point}
\acrodef{bs}[BS]{base station}
\acrodef{bse}[BSE]{beam squint effect}
\acrodef{cdf}[CDF]{cumulative distribution function}

\acrodef{coa}[COA]{curvature of arrival}
\acrodef{crb}[CRB]{Cram\'er-Rao bound}
\acrodef{ccrb}[CCRB]{constrained Cram\'er-Rao bound}
\acrodef{icrb}[ICRB]{intrinsic Cram\'er-Rao bound}
\acrodef{dof}[DOF]{degrees of freedom}

\acrodef{dbscan}[DBSCAN]{density-based spatial clustering of applications with noise}
\acrodef{elaa}[ELAA]{extremely-large antenna array}
\acrodef{ekf}[EKF]{extended Kalman filter}
\acrodef{ff}[FF]{far field}
\acrodef{fim}[FIM]{Fisher information matrix}
\acrodef{gnss}[GNSS]{global navigation satellite system}
\acrodef{gps}[GPS]{global positioning system}
\acrodef{icrb}[ICRB]{Intrinsic Cram\'er-Rao bound}
\acrodef{imu}[IMU]{inertial measurement unit}
\acrodef{ip}[IP]{incidence point}
\acrodef{kf}[KF]{Kalman Filter}
\acrodef{kld}[KLD]{Kullback–Leibler divergence}
\acrodef{las}[L\&S]{localization and sensing}
\acrodef{los}[LOS]{line-of-sight}
\acrodef{mae}[MAE]{mean absolute value}
\acrodef{map}[MAP]{maximum a posteriori}
\acrodef{mimo}[MIMO]{multiple-input-multiple-output}
\acrodef{mle}[MLE]{maximum likelihood estimator}
\acrodef{mm}[MM]{mismatched model}
\acrodef{mpc}[MPC]{multipath component}
\acrodef{nlos}[NLOS]{non-line-of-sight}
\acrodef{nf}[NF]{near field}
\acrodef{nr}[NR]{new radio}
\acrodef{ofdm}[OFDM]{orthogonal frequency division multiplexing}
\acrodef{pbd}[PBD]{partial blockage detection}
\acrodef{prs}[PRS]{positioning reference signal}
\acrodef{psd}[PSD]{power spectral density}
\acrodef{pss}[PSS]{primary synchronization signal}
\acrodef{rmse}[RMSE]{root mean squared error}
\acrodef{rf}[RF]{radio frequency}
\acrodef{rfc}[RFC]{radio frequency chain}

\acrodef{ris}[RIS]{reconfigurable intelligent surface}
\acrodef{rss}[RSS]{received signal strength}
\acrodef{rtk}[RTK]{real-time kinematic}
\acrodef{rtt}[RTT]{round-trip-time}
\acrodef{sa}[SA]{sub-array}
\acrodef{simo}[SIMO]{single-input-multiple-output}
\acrodef{slam}[SLAM]{simultaneous localization and mapping}
\acrodef{snr}[SNR]{signal-to-noise ratio}
\acrodef{sns}[SNS]{spatial non-stationarity}
\acrodef{sp}[SP]{scattering point}
\acrodef{ssb}[SSB]{synchronization signal/physical broadcast channel block}
\acrodef{swm}[SWM]{spherical wave model}
\acrodef{tdd}[TDD]{time division duplex}
\acrodef{tdoa}[TDOA]{time-difference-of-arrival}
\acrodef{tm}[TM]{true model}
\acrodef{toa}[TOA]{time-of-arrival}
\acrodef{ue}[UE]{user equipment}
\acrodef{upa}[UPA]{uniform planar array}
\acrodef{ura}[URA]{uniform rectangular array}
\acrodef{va}[VA]{virtual anchor}

\long\def\comment#1{}

\newfont{\bbb}{msbm10 scaled 700}

\newfont{\bb}{msbm10 scaled 1100}

\newcommand{\av}{{\bf a}}
\newcommand{\bv}{{\bf b}}
\newcommand{\cv}{{\bf c}}
\newcommand{\dv}{{\bf d}}
\newcommand{\ev}{{\bf e}}

\newcommand{\gv}{{\bf g}}
\newcommand{\hv}{{\bf h}}
\newcommand{\iv}{{\bf i}}

\newcommand{\mv}{{\bf m}}
\newcommand{\nv}{{\bf n}}

\newcommand{\pv}{{\bf p}}
\newcommand{\qv}{{\bf q}}
\newcommand{\rv}{{\bf r}}
\newcommand{\sv}{{\bf s}}
\newcommand{\tv}{{\bf t}}

\newcommand{\wv}{{\bf w}}
\newcommand{\vv}{{\bf v}}
\newcommand{\xv}{{\bf x}}

\newcommand{\zv}{{\bf z}}

\newcommand{\Am}{{\bf A}}
\newcommand{\Bm}{{\bf B}}

\newcommand{\Em}{{\bf E}}
\newcommand{\Fm}{{\bf F}}
\newcommand{\Gm}{{\bf G}}
\newcommand{\Hm}{{\bf H}}

\newcommand{\Jm}{{\bf J}}
\newcommand{\Km}{{\bf K}}

\newcommand{\Pm}{{\bf P}}
\newcommand{\Qm}{{\bf Q}}
\newcommand{\Rm}{{\bf R}}

\newcommand{\Tm}{{\bf T}}

\begin{document} 

\title{Intrinsic Cram\'er-Rao Bound based 6D Localization and Tracking for 5G/6G Systems}


\author{
Xueting~Xu,
Hui~Chen,~\IEEEmembership{Member,~IEEE},
Shengqiang~Shen,~\IEEEmembership{Member,~IEEE}, 
Hyowon~Kim,~\IEEEmembership{Member,~IEEE}, 
Xu~Fang,~\IEEEmembership{Member,~IEEE}, 
Ao~Peng,~\IEEEmembership{Member,~IEEE}, 
Fan~Jiang,~\IEEEmembership{Member,~IEEE}, 
and~Henk~Wymeersch,~\IEEEmembership{Fellow,~IEEE}

\thanks{X.~Xu and F. Jiang are with the Department of Broadband Communications, Pengcheng Laboratory, Shenzhen 518055, China (E-mail: xuxueting4728@gmail.com; jiangf02@pcl.ac.cn).}

\thanks{H. Chen and H.~Wymeersch are with the Department of Electrical Engineering, Chalmers University of Technology, 412 58 Gothenburg, Sweden (e-mail: {hui.chen; henkw}@chalmers.se).}

\thanks{S. Shen is with the School of Information and Control Engineering, China University of Mining and Technology, Xuzhou 221000, China
(e-mail: sshen@cumt.edu.cn).}

\thanks{Hyowon Kim is with the Department of Electronics Engineering, Chungnam National University, Daejeon 34134, South Korea (e-mail: hyowon.kim@cnu.ac.kr).}

\thanks{X. Fang is with the School of Electrical and Electronic Engineering, Nanyang Technological University, Singapore 639798 (e-mail: xu.fang@ntu.edu.sg).}

\thanks{A. Peng is with
the School of Informatics, Xiamen University, Xiamen, Fujian 361005,
China (e-mail: pa@xmu.edu.cn)}
}

\maketitle

\begin{abstract}
Localization and tracking are critical components of integrated sensing and communication (ISAC) systems, enhancing resource management, beamforming accuracy, and overall system reliability through precise sensing.
Due to the high path loss of the high-frequency systems, antenna arrays are required at the transmitter and receiver sides for beamforming gain. However, beam misalignment may occur, which requires accurate tracking of the six-dimensional (6D) state, namely, 3D position and 3D orientation. In this work, we first address the challenge that the rotation matrix, being part of the Lie group rather than Euclidean space, necessitates the derivation of the ICRB for an intrinsic performance benchmark. Then, leveraging the derived ICRB, we develop two filters-one utilizing pose fusion and the other employing error-state Kalman filter to estimate the UE's 6D state for different computational resource consumption and accuracy requirements. Simulation results validate the ICRB and assess the performance of the proposed filters, demonstrating their effectiveness and improved accuracy in 6D state tracking.

\end{abstract}


\IEEEpeerreviewmaketitle
\acresetall 

\section{Introduction}
\label{sec:intro}
Positioning is increasingly integrated into wireless network systems, and various applications that require both high-throughput communication and high-accuracy localization are emerging within the framework of Integrated Sensing and Communication (ISAC) systems \cite{chen2022tutorial}. For instance, virtual reality devices need real-time pose (position and orientation) tracking and video streaming for display~\cite{huang2023virtual}, while digital twinning relies on multi-modal sensing data from distributed infrastructure and user devices to make communication and sensing decisions~\cite{alkhateeb2023real}. Through ISAC, radio-based localization services can be seamlessly incorporated into existing communication infrastructures, complementing traditional techniques such as cameras, LiDAR, and others to provide enhanced sensing and communication capabilities.

Moving to the higher band (e.g., mmWave band) brings opportunities for bandwidth and array size that are suitable for the delay and angle estimation, respectively. 
When \ac{ue} is equipped with an array, orientation estimation can also be performed by the \ac{aoa} or \ac{aod} at the UE side.
In \cite{Shahmansoori2017position}, 2D position and 1D direction estimation Cram\'er-Rao Bound (CRB) is derived to evaluate the performance of position and orientation estimation, which is extended to 3D position and 2D direction in \cite{Guerra2018Single}. Based on this work, \cite{Abu2018Error} examines the performance bounds for uplink 5D estimation and analyzes the asymmetry between uplink and downlink performance. While 1D and 2D heading estimation for a moving UE are relatively easy, 3D orientation or 6D state (3D position and 3D orientation) requires manifold optimization due to the constraints of the rotation matrix~\cite{nazari2023mmwave}. In \cite{salem2025indoor}, 3D orientation is estimated during user motion, but 3D displacement is not considered simultaneously.
Several studies on 6D localization have been reported, with algorithms developed for scenarios such as multi-BS-aided scenario~\cite{zheng2023coverage} and multipath-aided single-BS localization~\cite{nazari2023mmwave}. 
However, these works only discuss position and orientation estimation at a single snapshot without considering the mobility of the target.

For moving devices, especially over long measurement durations, Doppler information can enhance channel parameter estimations. Alternatively, short localization snapshots ensure coherent channels, allowing a tracking algorithm to be applied across multiple snapshots. 
The 6D state modeling of mobile devices differs in the way rotations are represented and the choice of state space. 
Euler angles provide an intuitive way to represent orientation by clearly showing the rotation related to each axis~\cite{Talvitie2023orientation}, but they suffer from the fundamental problem of gimbal lock. Quaternions can solve this issue but, as a mathematical representation, their lack of direct connection to statistical estimation theory, combined with their computational complexity and non-intuitive nature, makes them a suboptimal choice for state tracking filters~\cite{huang2020global}.
Rotation matrices, accurately represent rotations, avoid the gimbal lock problem and have a physical meaning, making them well-suited for orientation tracking~\cite{suvorova2021tracking}.


Filters for state estimation can be defined over various domains or state spaces. In Euclidean space, common filters like the \ac{kf} and its variants \cite{Talvitie2023orientation, 10103759, 9893405} are widely used due to their effectiveness in linear and low nonlinear estimation problems. 
However, Euclidean-based filters struggle with rotation matrices and 6D states constrained on manifolds, as rotation matrices are closed under multiplication but not addition, making both the motion model and noise multiplicative. This can lead to a loss of orthogonality and estimation errors. Rotation matrices and 6D states can be effectively represented as compact Lie groups, which are closed under multiplication~\cite{varadarajan2013lie}. Lie algebras, associated with Lie groups, are closed under addition, enabling the linearization of multiplicative operations for simplifying 6D tracking. Gaussian distributions on Lie groups can be represented using concentrated Gaussian distributions~\cite{wang2006error}, enabling the use of filtering methods like the \ac{ekf} for Lie groups~\cite{bourmaud2015continuous}, via the exponential map from the Lie algebra to the Lie group~\cite{barfoot2014associating}, which is the core concept of the invariant \ac{ekf}~\cite{barrau2014intrinsic, tao2022adaptive}. While there are existing state tracking algorithms for Lie groups~\cite{xu2023distributed, zhang2022adaptive, petersen2022tracking}, the states and measurements they use differ from those in wireless signal-based 6D tracking, which remains an underexplored area.

In order to benchmark filters, it is also important to understand the fundamental performance bounds via Fisher information theory, tailored to the choice of orientation representation and state space. 
\ac{crb} is an effective tool to provide performance. Unlike position, which lies in Euclidean space, 3D orientation analysis has to consider the constraints on the rotation matrix. The \ac{ccrb}~\cite{stoica1998on} is an extension of the \ac{crb} for constrained parameters, utilizing Euclidean distance embedded within a Riemannian manifold for covariance calculation, making it applicable to rotation matrices. To better handle parameters constrained on a Riemannian manifold, \ac{icrb} uses the Riemannian or geodesic distance to calculate the covariance matrix of the parameters~\cite{smith2005covariance}, which is more natural for many applications. In addition, due to the orthogonality constraints, the \ac{dof} of a rotation matrix is fewer than the number of its elements. The CCRB, by bounding the RMSE of each element in a rotation matrix, ignores the relationships between elements, leading to a mismatch between the CCRB's order and the\ac{dof} of the rotation matrix~\cite{Nazari20213d}. Hence, the CCRB is unsuitable to represent the covariance of rotation matrices for tracking. In contrast, the ICRB addresses this issue by mapping parameters onto its tangent space, effectively reducing dimension~\cite{Boumal2013on} and ensuring that its order aligns with the \ac{dof}. This makes the ICRB an ideal tool for representing the covariance of rotation matrices. While the ICRB for state estimation is derived in \cite{labsir2024cramer} for observations in Euclidean space, estimating a rotation matrix using channel parameters is more complex since the observations—direction vectors—do not belong to Euclidean space.
To the best of our knowledge, the application of ICRB for estimating rotation matrices based on channel parameters remains unexplored so far.

This paper investigates the 6D localization of a connected device with the assistance of multiple (at least two) anchors (e.g., \acp{ap}). Instead of using CCRB, we derive the ICRB for 6D states estimation for 5G/6G systems, including a rotation matrix constrained on a manifold, and propose two ICRB-based 6D tracking filters to address nonlinear dynamics and multiplicative noise.
The main contributions of this work are summarized as follows.
\begin{itemize}
    \item \textbf{ICRB Characterization for MIMO-OFDM 6D estimation over manifolds:} We first derive the ICRB for channel geometric parameters constrained by the system, then extend this to compute the ICRB for 6D states by projecting these constrained parameters onto the tangent space of the manifold where the rotation matrix resides. This novel approach provides an intrinsic benchmark for 6D state estimation, allowing for characterizing the covariance matrix of rotation matrix measurements within the tracking filter.
    \item \textbf{6D tracking filters for constrained state using channel parameters:} We integrate the derived ICRB into two filtering methods—fusion-based and error-state EKF—to estimate a constrained 6D state, linking single-snapshot ICRB uncertainty to a multi-step estimation framework. Both filters leverage channel parameters and the derived ICRB to minimize estimation errors weighted by the covariance matrix, preventing errors from naive Euclidean approximations.
    \item \textbf{Simulation validation and performance analysis: } Simulation results demonstrate that the derived ICRB accurately captures the intrinsic error of the rotation matrix. Furthermore, the pose fusion-based algorithm shows superior estimation performance at the expense of computational resources, particularly in low SNR conditions. Both proposed 6D filters outperform the commonly used Euler angle-based tracking algorithm.
\end{itemize}

The structure of this paper is organized as follows. Section II discusses the system model and the background of Lie theory. The Fisher information analysis, including the ICRB for channel parameters and 6D state, is derived in Section III. 
Section IV details the proposed two 6D tracking filters based on the fusion method and error-state EKF, respectively. Simulation results are presented in Section V, followed by the conclusion of this work in Section VI.

\emph{Notations and Symbols:} Scalars are denoted by lowercase italic letters (e.g., $a$). Vectors are represented by lowercase boldface letters (e.g., $\mathbf{a}, \mathbf{w}$), while matrices are denoted by uppercase boldface letters (e.g., $\mathbf{A}$). $(\cdot)^\top$, $(\cdot)^{-1}$, $\text{tr}(\cdot)$, $\text{vec}(\cdot)$ and $\| \cdot \|_2$ indicate the transpose, inverse, trace, vectorization, and $\ell$-2 norm operations, respectively. $(\cdot)_{i }$ represents the $i$-th element of the vector. $(\cdot)_{i \cdot}$, $(\cdot)_{\cdot j}$, and $(\cdot)_{i j}$ represent the $i$-th row, the $j$-th column, and the element in the $i$-th row, $j$-th column of a matrix, respectively. $(\cdot)_{i:j,m:n}$ is the submatrix formed by the rows from $i$ to $j$ and the columns from $m$ to $n$ of a matrix. $\mathbb{R}^M$ is $M$-dimension real space and $\mathbb{S}^N$ is the unit sphere in $\mathbb{R}^{N+1}$. ${\widetilde {(\cdot)}}$ represents the parameter at the local coordinate system. The skew-symmetric operator $(\cdot)^{\wedge}$ denotes the conversion of a vector in $\mathbb{R}^3$ into a skew-symmetric matrix according to ${{\av}^ \wedge } = \left[ {0,-{a_3},{a_2};{a_3},0, - {a_1}; - {a_2},{a_1},0} \right] \in \mathbb{R}^{3 \times 3}$. $(\cdot)^\vee$ denotes the inverse of operator $(\cdot)^{\wedge}$, i.e., convert a skew-symmetric matrix in $\mathbb{R}^{3 \times 3}$ to a vector in $\mathbb{R}^3$. $(\cdot)^{\curlywedge} $ denotes the adjoint. $\text{SO}(n)$ and $\text{SE}(n)$ denote $n$-dimension special orthogonal group and $n$-dimension special European group. $\mathfrak{so}(n)$ and $\mathfrak{se}(n)$ denote their corresponding Lie algebras. $\text{grad}(\cdot)$ denotes the gradient. For a function, $\overline{(\cdot)}$ denotes the smooth extension; for a state, $\overline{(\cdot)}$ denotes the noise-free nominal value. $(\cdot)^{-}$ and $(\cdot)^{+}$ denote the prior estimate and posterior estimate. $\hat{(\cdot)}$ denotes the measurement.



\section{System Model}
In this section, we start with the definition of the signal model, channel parameters, and UE states. Then, the background of the Lie theory is described as a mathematical basis for 6D filter design in the subsequent sections.

\subsection{Signal Model}
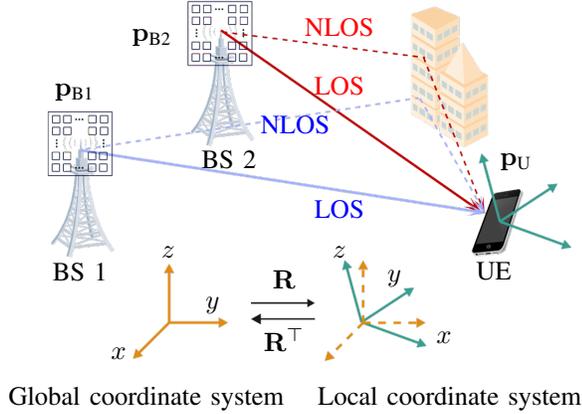
\begin{figure}
    \centering
    \begin{tikzpicture}
    \node (image) [anchor=south west]{\includegraphics[width=0.8\linewidth]{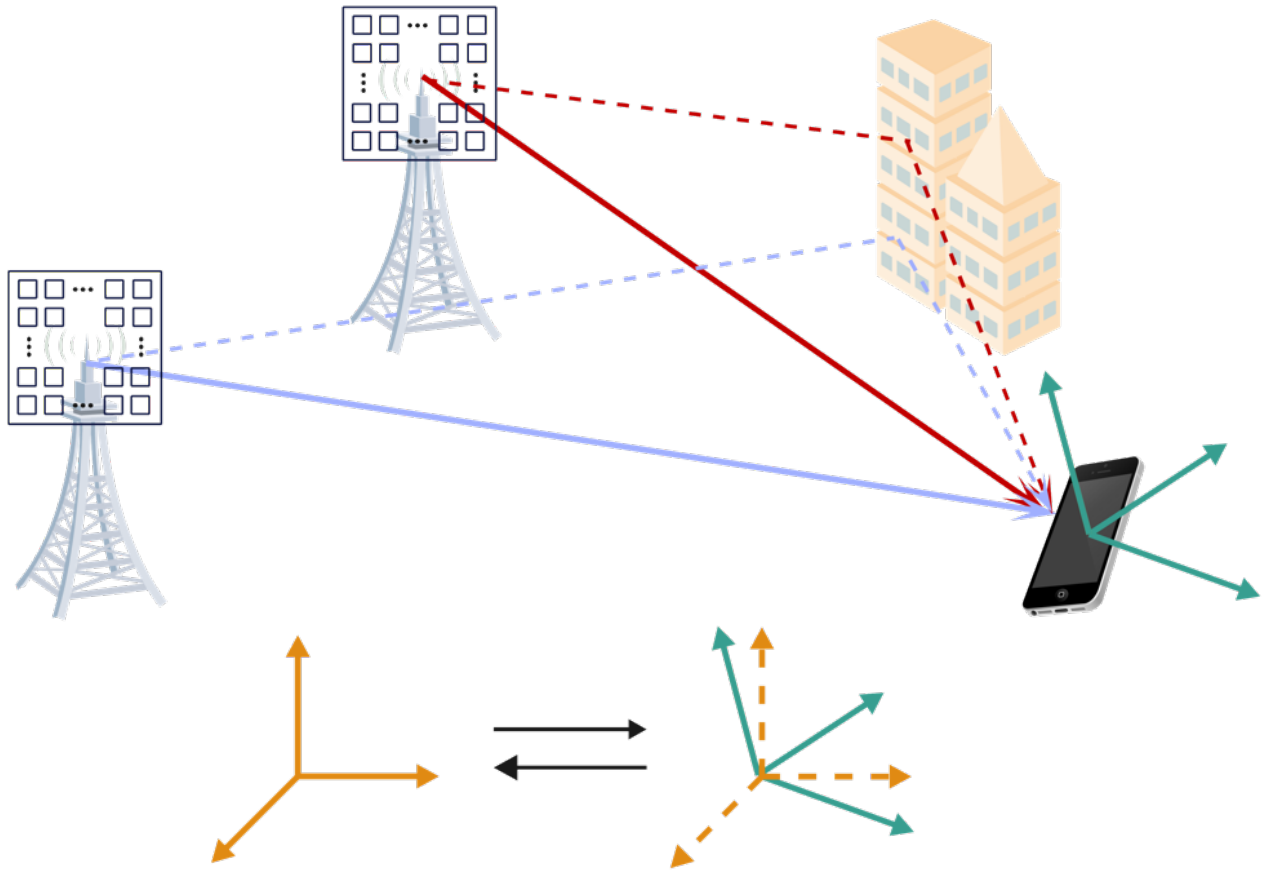}};
    \gettikzxy{(image.north east)}{\ix}{\iy};
    \node at (0.21*\ix,0.92*\iy)[rotate=0,anchor=north]{{${\pv_{\text{B2}}}$ }};
    \node at (0.07*\ix,0.78*\iy)[rotate=0,anchor=north]{{${\pv_{\text{B1}}}$ }};
    \node at (0.87*\ix,0.6*\iy)[rotate=0,anchor=north]{{$\pv_{\text{U}}$ }};
    \node at (0.24*\ix,0.36*\iy)[rotate=0,anchor=north]{{$z$}};
    \node at (0.55*\ix,0.36*\iy)[rotate=0,anchor=north]{{$z$}};
    \node at (0.15*\ix,0.1*\iy)[rotate=0,anchor=north]{{$x$}};
    \node at (0.74*\ix,0.14*\iy)[rotate=0,anchor=north]{{$x$}};
    \node at (0.32*\ix,0.23*\iy)[rotate=0,anchor=north]{{$y$}};
    \node at (0.65*\ix,0.3*\iy)[rotate=0,anchor=north]{{$y$}};
    \node at (0.45*\ix,0.15*\iy)[rotate=0,anchor=north]{{$\Rm^\top$}};
    \node at (0.45*\ix,0.3*\iy)[rotate=0,anchor=north]{{$\Rm$}};
    \node at (0.08*\ix,0.32*\iy)[rotate=0,anchor=north]{{BS 1}};
    \node at (0.35*\ix,0.61*\iy)[rotate=0,anchor=north]{{BS 2}};
    \node at (0.83*\ix,0.32*\iy)[rotate=0,anchor=north]{{UE}};
    \node at (0.55*\ix,0.95*\iy)[rotate=0,anchor=north]{{\red{NLOS}}};
    \node at (0.55*\ix,0.8*\iy)[rotate=0,anchor=north]{{\red{LOS}}};
    \node at (0.47*\ix,0.7*\iy)[rotate=0,anchor=north]{{\blue{NLOS}}};
    \node at (0.55*\ix,0.48*\iy)[rotate=0,anchor=north]{{\blue{LOS}}};
    \node at (0.2*\ix,0*\iy)[rotate=0,anchor=north]{{Global coordinate system}};
    \node at (0.75*\ix,0*\iy)[rotate=0,anchor=north]{{Local coordinate system}};
    \end{tikzpicture}
    \caption{An example of a 3D MIMO system scenario. The system includes two BSs equipped with large-scale antenna arrays and one UE. The rotation of the UE $\Rm$ results in a difference between the global and local coordinate systems, which can be represented by $\Rm$. The system includes both LOS and NLOS paths, which can be distinguished. We use the LOS paths for localization.}
    \label{scenario}
\end{figure}

We consider a downlink 3D MIMO-OFDM system consisting of $N > 1$ BSs and a UE. We will focus on $N=2$ without loss of generality. The BSs are anchors with known positions ($\pv_{\text{B1}},\pv_{\text{B2}} \in \mathbb{R}^{3}$) and orientations ($\Rm_{\text{B1}}, \Rm_{\text{B2}} \in \text{SO(3)}$, which are rotation matrix representing the orientation of the BSs). The position $\pv_{\text{U},k}$ and orientation $\Rm_{\text{U},k}$ of the UE are both unknown. By assuming orthogonal transmissions of two BSs (e.g., in time or frequency), we take the channel of one BS-UE pair for example. The BS and the UE respectively equipped with ${{\text{N}_{\text{B}}}}$ and ${{\text{N}_{\text{U}}}}$ radiation elements forming \acp{upa}. In the OFDM system, the channel parameters are approximately constant for a short time. Assuming a downlink scenario at the discrete time $k$, the received signal of UE at the $g$-th transmission and $c$-th subcarrier can be expressed as\footnote{The downlink signal model applies for different BSs. By transmitting time- or frequency-orthogonal signals, the UE can resolve the signals from two BSs to estimate the channel geometric parameter information contained in each signal.}

\begin{equation}
    y_{g,c,k} = \bv_{\text{U},g,k}^\top \Hm_{c,k} \bv_{\text{B},g,k} x_{c,g,k} + n_{c,g,k},
    \label{2-1}
\end{equation}
where $\bv_{\text{B},g,k}$ is the precoding vector at the BS, $\bv_{\text{U},g,k}$ is the combining vector at the UE, $x_{c,g,k}$ is the transmit signal before the precoder, and $n_{c,g,k} \sim \mathcal{CN}(0,\sigma_k^2)$ is the complex additive white Gaussian noise. The considered system model is shown in Fig.~\ref{scenario}.

According to the relationship between coherent time and maximum Doppler frequency, we can ignore the Doppler effect in a snapshot with a short time interval~\cite{wymeersch2022radio}~\footnote{The Doppler effects can generally be ignored when the transmission time in localization satisfy $\|K T_s f_D\| \ll 1 $ \cite{wymeersch2022radio}, where $K$ is the number of beams, $T_s$ is the symbol duration, and $f_D$ is the Doppler shift. Suppose we have $K = 20$ beam pairs and $T_s = 20$ ns. The Doppler shift can be calculated by $f_D = \frac{v}{c} f_c$, where $f_c$ is the carrier frequency and $v$ is relative radial velocity. In the case of $f_c = 30$ GHz, when $v \ll 50$ m/s, the Doppler effect can be ignored. This velocity works for most mobile equipment (e.g., 5 m/s is sufficient for most indoor applications).}. Hence, the frequency-flat fading channel matrix at subcarrier $c$ can be expressed as
\begin{equation}
\begin{aligned}
\Hm_{c,k} &= {\underbrace {\alpha_{k} \av_{\text{U}}(\tv_{\text{UB},k}) \av_{\text{B}}^\top (\tv_{\text{BU},k}) e^{-j2\pi\left[\tau_{k}(c-1)\Delta_f\right]}}_{\text{LOS path}} }\\
& + {\underbrace { \sum\limits_{p = 1}^P  {\alpha _{p,k}}{\av_{\text{U}}}({\tv_{p,\text{UB},k}}) \av_{\text{B}} ^ \top ({\tv_{p, \text{BU},k}}){e^{ - j2\pi \left[ {{\tau _{p, k}}(c - 1){\Delta _f}} \right]}}}_{\text{NLOS paths}}} 
\label{2-3b}
\end{aligned}
\end{equation}
where for the \ac{los} path, $\alpha_{k}$, $\av_{\text{U}}$, $\av_{\text{B}}$, $\tv$, $\tau_k$ are the complex channel gain (assumed to be identical for different subcarriers), steering vector at the UE, steering vector at the BS, direction vector, and delay at time $k$, respectively. The parameters for the \ac{nlos} paths~\footnote{For the \ac{nlos} path, each path may involve single or multi-bounce reflections or diffuse scattering, which is beyond the main focus of this work. Therefore, the NLOS path is excluded from UE localization in this work, by assuming the LOS path can be resolved from the NLOS paths due to the large bandwidth and antenna aperture size of the system \cite{Abu2018Error}.} are defined in a similar way, with $p$ indicating the $p$-th path, and $\Delta_f$ is the subcarrier spacing. For notation brevity, we omit the time index $k$ in the following description. 

The local direction vector $\tv \in \mathbb{S}^2$ is constrained in a unit circle, that is, $\tv$ is a $3\times1$ unit norm vector. As for the direction of $\tv$ (we take $\tv_\text{BU}$ as an example), 
\begin{equation}
\tv_\text{BU} = {\bf{R}}_\text{B}^{-1} \frac{\xv_\text{U}-\xv_\text{B}} {\Vert \xv_\text{U}-\xv_\text{B} \Vert} ,
\label{9}
\end{equation}
and ${\bf{R}}_\text{B} \tv_\text{BU} = -{\bf{R}}_\text{U} \tv_\text{UB}$. 
For simplicity, let $\text{D}\in \{{\text{U, B}}\}$ denote the device type where antennas are mounted, where U and B denote the UE and BS, respectively. We assume the 3D geometric centers of antenna arrays at the device is located at $\pv_{\rm{D}}$ in the global coordinate system. The signal delay $\tau$ is composed of propagation delay and clock offset, which is defined as: 
\begin{equation}
\begin{aligned}
{ \tau } &= \frac{1}{c_l}\left\| {{\pv_{\text{U}}} - {\pv_{\text{B}}}} \right\|+ B,
\label{tau}
\end{aligned}
\end{equation}
where $c_l$ is the speed of light and $B$ is the synchronization offset of the UE. The response vector of a specific antenna array $\av_{\rm{D}}(\tv)$ is defined as
\begin{equation}
{{\av}_{\rm{D}}}\left( \tv \right) = {\left[ {{a_{\text{D},1}}, \ldots ,{a_{\text{D},d}}, \ldots ,{a_{\text{D},{{\text{N}_{\rm{D}}}}}}} \right]^\top}, {a_{\text{D},d}} = e^{j\frac{{2\pi \blue{f_c}}}{c_l} {{{{\pv^\top_d}}} \tv } },
\end{equation}
where $f_c$ is the carrier frequency, and $\pv_d$ is 3D local position of the $d$-th element on the array.


\subsection{Channel Parameters and UE States}\label{2-3}
We define the channel parameter vector at time $k$ as $ \zv_k= \left[ {{ \tau _{\text{B},k}},{ {\tv} _{\text{UB},k}}^\top, { {\tv} _{\text{BU},k}}^\top }, \mathfrak{R}(\alpha_{\text{B},k}), \mathfrak{I}(\alpha_{\text{B},k}) 
\right]^\top $, which can be obtained by standard channel parameter estimation techniques, such as estimating signal parameters via rotational invariance techniques (ESPRIT) \cite{keskin2023esprit}, multiple signal classification (MUSIC) \cite{9965430}, compressive sensing \cite{10273424}, etc. The UE state comprises the 3D positions and 3D orientations, and we adopt the Lie theory \cite{humphreys2012introduction}, a powerful tool to model either displacement or transformation of the 3D points with 3D orientations. With the Lie theory, we denote the UE state at time $k$ as
\begin{equation}
    {\Tm_{\text{U},k}} \buildrel \Delta \over = \left[ {\begin{array}{*{20}{c}}
{\Rm_{\text{U},k}}&{{\Rm_{\text{U},k}} {\pv_{\text{U},k}}}\\
{\bf{0}}_{3}^\top&1
\end{array}} \right] \in \mathbb{R}^{4 \times 4}
\label{state}
\end{equation}
Equation \eqref{state} represents a transformation in a Lie group combining rotations and positions in 3D, showing how they are expressed and manipulated within the group structure.
We can estimate the position and orientation of UE at a single snapshot if there are at least 2 BSs in LOS using the relationship between the channel parameter vector and UE state vector according to \eqref{9} and \eqref{tau}. 

For the dynamics of moving UE over a period of time, we adopt the linear constant-velocity model for both rotation matrix and position \cite{legnani1996homogeneous}:
\begin{subequations}
\begin{align}
\Rm_{{\text{U}},k} &= \text{exp} (\Delta _k \wv_k  ^ \wedge) \Rm_{{\text{U}},k-1},\label{4-2}\\
\pv_{{\text{U}},k} &= \Rm_{{\text{U}},k}^\top \Delta _k \vv_k + \pv_{{\text{U}},k-1}, \label{4-2a}
\end{align}
\end{subequations}
where $\Delta _k$ is the time interval of two adjacent sample times, $\vv_k$ is the translation velocity at the local coordinate system, $\wv_k$ is the rotational velocity, and the subscript $k$ indicates the tracking epoch. By combining the kinematics of rotation and position, we can get the kinematics of the UE state as:
\begin{equation}
\begin{aligned}
{\Tm_{{\text{U}},k}} = \left[ {\begin{array}{*{20}{c}}
{\text{exp} (\Delta _k \wv_k  ^ \wedge)}&{\Delta _k \vv_k}\\
{{\bf{0}}_{3}}^\top&1
\end{array}} \right] {\Tm_{{\text{U}},k - 1}},
\end{aligned}
\label{kine}
\end{equation}


Therefore, we have defined the channel parameter vector $\zv_k$, UE state ${\Tm_{\text{U},k}}$ and its kinematic model. In the following sections, we derive the estimation performance bound of channel parameters with constraints, as well as the corresponding bound of the unknown state accordingly. Based on the derived bound and measurements, we propose 6D filters to predict and update the state of the moving UE continuously.


\subsection{Background on Lie Theory}


In the following content, we will frequently use Lie theory, including using it for the definition of the UE state. We provide some background knowledge of Lie theory here.

Rotation matrix ${\Rm_{\text{U}}}$ and UE state matrix ${\Tm_{\text{U}}}$ belong to ${\text{SO}(3)}$ and ${\text{SO}(3)}$, respectively, whose definition are \cite[(7.1), (7.5)]{barfoot2024state}:
\begin{equation}
\begin{aligned}
& \text{SO}(3)=\{\Rm \in \mathbb{R}^{3 \times 3} | \Rm^\top \Rm = {\bf{I}} , \text{det}(\Rm)=1\} \\
& \text{SE}(3)=\left\{\left. \Tm =\left[\begin{array}{cc}
\Rm & \bv \\
\mathbf{0}^\top & 1
\end{array}\right] \in \mathbb{R}^{4 \times 4 } \right\rvert\, \Rm \in \text{SO}(3), \bv \in \mathbb{R}^3 \right\}
\end{aligned}
\end{equation}
$\text{SO}(3)$ and $\text{SE}(3)$ are both Lie groups, which are also a smooth manifold, with group operations (multiplication and inversion) that are smooth maps. It can be seen that the $\text{SO}(3)$ and $\text{SE}(3)$ are constrained by orthogonality, which makes many operations, such as addition, no longer closed for these Lie groups. 

Every matrix Lie group is associated with a Lie algebra, with $\mathfrak{so}(3)$ and $\mathfrak{se}(3)$ as the Lie algebras for $\text{SO}(3)$ and $\text{SE}(3)$, respectively. A Lie algebra is a vector space that serves as the tangent space of the Lie group at its identity element,
capturing the local structure of the corresponding group \cite{barfoot2024state}. Lie algebras are closed under addition, so they can assist in performing some operations on Lie groups, making them useful for tasks like UE state estimation.

\subsubsection{{\text{SO}}(3)}
The conversion between Lie group and Lie algebra can be realized by the so-called exponential map and logarithmic map. We first give the mapping between $\text{SO}(3)$ and its Lie algebra $\mathfrak{so}(n)$, which is useful for the operation of the rotation matrix. The Lie algebra element corresponding to Lie group element $\Rm \in \text{SO}(3)$ is $\rv ^\wedge \in \mathfrak{so}(3)$. If we have a known $\rv \in \mathbb{R}^{3} $, we can first decompose it as $\rv = \lambda {\bm{\zeta}}$, where $\lambda$ is the norm of $\rv$ and $\bm{\zeta}$ is a unit direction vector. Then we can obtain the corresponding rotation matrix by Rodrigues formula as \cite{dai2015euler}: 
\begin{equation}
    \Rm = \text{exp}(\rv ^\wedge) = \text{cos}\lambda \textbf{I} +(1-\text{cos}\lambda) {\bm{\zeta}} {\bm{\zeta}}^\top + \text{sin}\lambda {\bm{\zeta}}^{\wedge},
    \label{3-1-2}
\end{equation}
where $\text{exp}(\cdot)$ is the matrix exponential function. Given a square matrix $\Am$, the matrix exponential $\text{exp} (\Am) $ is defined by the power series \cite{cardoso2010exponentials}: $\text{exp} (\Am) = \sum\nolimits_{i = 0}^\infty ({\Am^i}/{i!})$ where $i!$ is the factorial of $i$. It can be shown that $\rv$ is an axis-angle presentation of $\Rm$, which describes a rotation in three-dimensional space by specifying an axis of rotation and an angle of rotation around that axis. 

If we have a known $\Rm$, according to Euler’s rotation theorem, $\rv$ can be calculated by the logarithmic map as \cite{shen2020simultaneous}
\begin{equation}
    \rv = [\text{log}(\Rm)]^\vee  = \frac{\lambda}{2\sin(\lambda)}
    \begin{bmatrix}
        [\Rm]_{3,2} - [\Rm]_{2,3}\\
        [\Rm]_{1,3} - [\Rm]_{3,1}\\
        [\Rm]_{2,1} - [\Rm]_{1,2}
    \end{bmatrix},
    \label{3-1-1}
\end{equation}
where rotation angle $\lambda = \arccos{\left\{ \left[ {\text{tr}(\Rm)-1} \right]/2 \right\} }$, and $\text{log}(\cdot)$ is the matrix logarithm function.Given a square matrix $\Am$, the matrix logarithm $\text{log} (\Am) $ is a matrix $\Bm$ such that $\text{exp}(\Bm) = \Am$ \cite{cardoso2010exponentials}.
\subsubsection{\text{SE}(3)}
For Lie group element $\Tm \in \text{SE}(3)$, whose corresponding Lie algebra element is $\bm{\xi}^{\wedge} = ([\bm{{\rho}}^\top, \rv^\top]^\top)^{\wedge} \in \mathfrak{se}(3)$, the skew operator is overloaded for $\text{SE}(3)$ as \cite[(7.14)]{barfoot2024state}:
\begin{equation}
    \bm{\xi}^{\wedge}= \left[ {\begin{array}{*{20}{c}}
{\bm{{\rho}}}\\ \rv
\end{array}} \right]^{\wedge}  = \left[ {\begin{array}{*{20}{c}}
{\rv ^ \wedge}&{{{\bm{{\rho}}}} }\\
{\bm{0}^\top}&{0}
\end{array}} \right] \in \mathbb{R}^{4\times4}.
\label{se3skew}
\end{equation}
The exponential map includes the map of $\pv$ in addition to $\Rm$, which is written as \cite[(7.33), (7.34)]{barfoot2024state}:
\begin{equation}
    \Tm = \text{exp} \left( \bm{\xi}^{\wedge} \right) = \left[ {\begin{array}{*{20}{c}}
{\text{exp}(\rv^\wedge)}&{{\Jm}_l{\bm{\rho}}}\\
{{\bm{0}^\top}}& 1
\end{array}} \right],
\label{3-1-4}
\end{equation}
where $\Jm_l \in \mathbb{R}^{3\times3}$ is the left Jacobian matrix of $\text{SO}(3)$ given in closed form as \cite[(7.37a)]{barfoot2024state}:
\begin{equation}
    \Jm_l = \frac{{\sin \lambda }}{\lambda } \textbf{I} + \left( {1 - \frac{{\sin \lambda }}{\lambda }} \right){\bm{\zeta}} {{\bm{\zeta}} ^T} + \frac{{1 - \cos \lambda }}{\lambda }{{\bm{\zeta}} ^ \wedge }.
\label{3-1-5}
\end{equation}
In the following, we omit the subscript $l$ and use $\Jm$ to represent the left Jacobian matrix. In the logarithmic map of $\text{SE}(3)$, ${\rv}$ can be calculated by rotation matrix $\Rm$. ${\bm{\rho}}$ can be solved by the linear equation $\pv = {\Jm{\bm{\rho}}}$.

\subsubsection{Adjoint representation}
The Lie group homomorphism is defined by the mapping $\Psi(\Tm) \Gm = \Tm \Gm \Tm^{-1}$, where $\Tm \in \text{SE}(3)$ and $\Gm \in \text{SE}(3)$. The adjoint representation of $\Tm$ is defined as the derivative of $\Psi(\Tm)$ at the origin. For a small Lie algebra element in $\mathfrak{se}(3)$, the adjoint representation can be simplified as the following linear operator:
\begin{equation}
\text{Ad}(\Tm) \mathfrak{g}^{\wedge} =\Tm \mathfrak{g}^{\wedge} \Tm^{-1},
\label{original_Ad}
\end{equation}
where $\mathfrak{g}^{\wedge}$ is a Lie algebra element in $\mathfrak{se}(3)$.

By differentiating at the identity element, the adjoint representation of a Lie group can be extended to a representation of its Lie algebra, which can also be calculated by the Lie bracket of vector fields as follow \cite[(7.16), (7.17)]{barfoot2024state}:
\begin{equation}
    \bm{\xi}^{\curlywedge} = \text{ad}(\bm{\xi}^{\wedge}) = \left[ {\begin{array}{*{20}{c}}
{\bm{{\rho}}}\\ {\rv}
\end{array}} \right]^{\curlywedge}  = \left[ {\begin{array}{*{20}{c}}
{\rv ^ \wedge}&{{{\bm{{\rho}}}} ^ \wedge}\\
{\bm{0}}&{\rv ^ \wedge}
\end{array}} \right] \in \mathbb{R}^{6\times6}.
\label{Ad}
\end{equation}
The mapping from $\text{ad}(\bm{\xi}^{\wedge})$ to $\text{Ad}(\Tm)$ is an exponential map, so $\text{Ad}(\Tm)$ can be expressed as \cite[(7.59)]{barfoot2024state}: 
\begin{equation}
    \text{Ad}(\Tm) = \text{Ad}\left(\left[\begin{array}{ll}
\Rm & {{\Jm}_l{\bm{\rho}}} \\
\mathbf{0}^\top & 1
\end{array}\right]\right)=\left[\begin{array}{cc}
\Rm & ({{\Jm}_l{\bm{\rho}}})^{\wedge} \Rm \\
\mathbf{0} & \Rm
\end{array}\right] \in \mathbb{R}^{6\times6}.
\label{Ad(SE3)}
\end{equation}
Note, the uppercase $\text{Ad}(\cdot)$ represents the adjoint matrix of $\text{SE}(3)$; for $\mathfrak{se}(3)$, the lowercase notation $\text{ad}(\cdot)$ is used. $\text{Ad}(\cdot)$ is the exponential map of $\text{ad}(\cdot)$.
\subsubsection{Baker–Campbell–Hausdorff (BCH) Formula}
While the mapping between the Lie group and Lie algebra can be realized using exponential and logarithmic functions, it's important to note the exponential multiplication property of scalars does not hold for matrices. In other words, the product of two matrices on a Lie group does not correspond to the sum of two vectors on a Lie algebra. The product of matrix exponents can be computed by the BCH formula \cite{barfoot2024state}. For $\text{SE}(3)$, the approximate expression of the BCH formula is \cite[(7.83a)]{barfoot2024state} :
\begin{equation}
\begin{aligned}
    \text{log}\left( \Tm_1 \Tm_2 \right)^\vee &= \text{log}\left[ \text{exp} \left( \bm{\xi}_1^{\wedge} \right) \text{exp} \left( \bm{\xi}_2^{\wedge} \right)  \right]^\vee \\
    &\approx \left\{ {\begin{array}{*{20}{c}}
{\boldsymbol{\mathcal{J}}}_l\left(\bm{\xi}_2\right)^{-1}{\bm{\xi}_1} + {\bm{\xi}_2}  \text{\; , if \;} {\bm{\xi}_1} \text{ \; small}\\
{\bm{\xi}_1} + {\boldsymbol{\mathcal{J}}}_r\left(\bm{\xi}_1\right)^{-1} \bm{\xi}_2   \text{\; , if \;} {\bm{\xi}_2} \text{ \; small}
\end{array}} \right.
\end{aligned}
\label{3-1-7}
\end{equation}
where $\bm{\xi}_1$ and $\bm{\xi}_2$ are the Lie algebra elements of $\Tm_1$ and $\Tm_2$. $\boldsymbol{\mathcal{J}}_l \in \mathbb{R}^{6\times6} $ is the left Jacobian matrix of $\text{SE}(3)$, which can be approximately calculated by \cite[(7.91)]{barfoot2024state}
\begin{equation}
    \boldsymbol{\mathcal{J}}_l = \sum \limits_{n = 0}^\infty \frac{1}{(n+1)!} {(\bm{\xi}^{\curlywedge})}^n 
    \approx \textbf{I}_6 + \frac{1}{2} \bm{\xi}^{\curlywedge},
    \label{3-1-6}
\end{equation}
${\boldsymbol{\mathcal{J}}}_r$ is the right Jacobian matrix derived by \cite[(7.80)]{barfoot2024state}
\begin{equation}
 {\boldsymbol{\mathcal{J}}}_r\left(\bm{\xi}\right) = {\boldsymbol{\mathcal{J}}}_l\left(-\bm{\xi}\right).
\label{3-1-8}
\end{equation}

Based on the described system model and the Lie theory background, we can proceed with Fisher information analysis and 6D filter design, as detailed below.

\section{Fisher Information Analysis}
In this section, we derive the performance bound of the UE states using a two-stage approach. In the first stage, we convert IQ-samples into channel parameters and derive the performance bounds of channel parameters. Based on the bound, we convert the channel parameters into the UE states and derive the performance bounds of the UE states in the second stage. This structure ensures efficient derivation and understanding of performance bound of the UE state with constraints. The ICRB suitable for the constrained state is employed for determining both bounds.

\subsection{Background on Fisher Information}
\begin{figure}
    \centering
    \begin{tikzpicture}
    \node (image) [anchor=south west]{\includegraphics[width=0.4\linewidth]{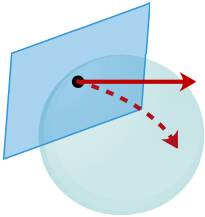}};
    \gettikzxy{(image.north east)}{\ix}{\iy};
    \node at (0.5*\ix,0.85*\iy)[rotate=0,anchor=north]{{$\Tm_\zv {\mathbb{Q}}$}};
    \node at (1.05*\ix,0.8*\iy)[rotate=0,anchor=north]{{${{\text{Log} }_\zv }(\hat \zv ) \in \Tm_\zv {\mathbb{Q}}$}};
    \node at (0.55*\ix,0.5*\iy)[rotate=0,anchor=north]{{$\dv_{\zv \rightarrow {\hat \zv}}$}};
    \node at (0.3*\ix,0.6*\iy)[rotate=0,anchor=north]{{$\zv$}};
    \node at (0.82*\ix,0.35*\iy)[rotate=0,anchor=north]{{$\hat \zv$}};
    \node at (0.5*\ix,0.2*\iy)[rotate=0,anchor=north]{{$ {\mathbb{Q}}$}};
    \end{tikzpicture}
    \caption{ Diagram of the logarithmic map. The geodesic distance $\parallel \dv_{\zv \rightarrow {\hat \zv}} \parallel_2$ between two points $\zv$ and $\hat \zv $ on the manifold is the distance between point $\zv$ and the projection of $\hat \zv $ onto the tangent space at $\zv$, that is, $\parallel \dv_{\zv \rightarrow {\hat \zv}} \parallel_2 = \parallel {{\text{Log} }_\zv }(\hat \zv ) \parallel_2 $.
    }
    \label{geodesic}
\end{figure}

Error bounds are commonly used as an effective performance evaluation tool for geometry-based localization methods. Different performance bounds are selected based on parameter constraints. For instance, the position in Euclidean space can be bounded by CRB~\cite{kay1993fundamentals}, and the rotation matrix can be lower bounded by CCRB~\cite{stoica1998on}. Next, we briefly describe ICRB and the derivation for 6D positioning.
Unlike the CCRB, the covariance matrix related to the \ac{icrb} is formulated regarding Riemannian distances on the manifold, providing a more accurate characterization of parameters within the Riemannian manifold. For a parameter $\zv$ constrained on manifold ${\mathbb{Q}}$, the ICRB related to covariance matrix is defined by \cite[(5)]{Boumal2013on}:
\begin{equation}
    \begin{aligned}
       & \mathbb{E} \left\{ {{{\left\langle {{{\text{Log} }_\zv }(\hat \zv ),{{\ev}_i}} \right\rangle }_\zv } \cdot {{\left\langle {{{\text{Log} }_\zv }(\hat \zv ),{{\ev}_j}} \right\rangle }_\zv }} \right\} \\
       & \succeq [{\Fm^{-1}_{\zv, \text{ICRB}}}]_{i,j}  \ + \ \text{curvature} \ \text{term}, 
        \label{3-2-5}
    \end{aligned}
\end{equation}
where ${\text{Log} }_\zv : {\mathbb{Q}} \rightarrow \Tm_\zv {\mathbb{Q}}$ means the logarithmic map from a point of the manifold ${\mathbb{Q}}$ to a point in the tangent space $\Tm_\zv {\mathbb{Q}}$. The tangent space $\Tm_\zv {\mathbb{Q}}$ is a vector space consisting of all possible directions in which one can tangentially pass through $\zv$ on the manifold ${\mathbb{Q}}$. By using the tangent space for error characterization, we simplify computations while maintaining the geodesic distance and intrinsic geometry of the manifold, as shown in Fig. \ref{geodesic}. $\ev_i$ and $\ev_j$ are the orthonormal basis of the tangent space $\Tm_\zv {\mathbb{Q}}$. Metric $\langle \cdot, \cdot \rangle_\zv$ is Riemannian metric on a smooth manifold ${\mathbb{Q}}$, which is a smoothly varying positive-definite inner product on the tangent spaces, defined as the infimum of the lengths of all smooth curves connecting the two points, i.e., geodesic distance. The curvature terms involve the manifold's sectional and Riemannian curvatures, vanishing if manifold ${\mathbb{Q}}$ is flat (locally Euclidean). For non-flat manifolds, these terms can often be neglected at high SNR.

The corresponding FIM is defined as \cite[(21)]{Boumal2013on}:
\begin{equation}
\begin{aligned}
 {\left[ {\Fm_{\zv, \text{ICRB}}} \right]_{i,j}}  & = \mathbb{E}\left\{ {{\rm{D}}{\cal L}(\zv)\left[ {{{\bf{e}}_i}} \right] \cdot {\rm{D}}{\cal L}(\zv)\left[ {{{\bf{e}}_j}} \right]} \right\} 
 \label{5-37}
\end{aligned}
\end{equation}
where ${\cal L}(\zv)$ denotes the log-likelihood function of $\zv$, and ${\rm{D}}{\cal L}(\zv)\left[ {{{\bf{e}}_i}} \right]$ denotes the Riemannian differential. 
${\rm{D}}{\cal L}(\zv)\left[ {{{\bf{e}}_i}} \right] = {\rm{d}}\{ {\cal L} [c(t)] \}/\rm{d}t|_{t=0}$ is the Riemannian differential of a smooth scalar field $\mathbb{Q}$ at a point $\zv$ along a tangent vector ${\bf{e}}_i$, where $c$ is a smooth curve on manifold passing through $\zv$ at $t=0$ with velocity ${\bf{e}}_i$ \cite[(3.27)]{barfoot2024state}. Riemannian differential gives the change in the function along any given direction.

The Riemannian differential can be calculated by Riemannian gradient, which is defined as:
\begin{equation}
\begin{aligned}
\left\langle { {{\rm{grad}}{\cal L}} ({\zv}),{\ev_i}}\right\rangle_{\zv} = {\rm{D}}{\cal L}(\zv)\left[ {{{\bf{e}}_i}} \right] 
 \label{grad}
\end{aligned}
\end{equation}
where ${{\rm{grad}}{\cal L}(\zv)}$ denotes the Riemannian gradient of ${\cal L}(\zv)$. 
The Riemannian gradient is a vector field that represents the rate of change in the direction of the steepest ascent or descent of a smooth function on the tangent space of the Riemannian manifold. Therefore, the Riemannian gradient is an orthogonal projection of the smoothly extended Euclidean gradient to the tangent space \cite{boumal2023introduction}, which is given by 
\begin{equation}
{\rm{grad}}{\cal L}\left( {\zv} \right) = {{\rm{Proj}}_{\zv}}\left[ {{\rm{grad}}\overline {\cal L} \left( {\zv} \right)} \right] \in \mathbb{R} ,
\label{3-5}
\end{equation}
\begin{figure}
    \centering
    \begin{tikzpicture}
    \node (image) [anchor=south west]{\includegraphics[width=0.4\linewidth]{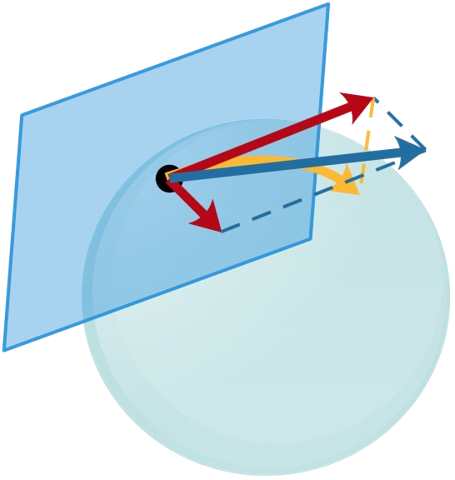}};
    \gettikzxy{(image.north east)}{\ix}{\iy};
    \node at (1.25*\ix,0.9*\iy)[rotate=0,anchor=north]{{$\red{{\rm{grad}}{\cal L}\left( {\bf{t}} \right) = {\rm{grad}} \overline {\cal L}\left( {\bf{t}} \right)_\parallel}$ }};
    \node at (1.1*\ix,0.75*\iy)[rotate=0,anchor=north]{{$\blue{{\rm{grad}} \overline {\cal L}\left( {\bf{t}} \right)}$}};
    \node at (0.55*\ix,0.5*\iy)[rotate=0,anchor=north]{{$\red{{\rm{grad}} \overline {\cal L}\left( {\bf{t}} \right)_\bot}$}};
    \node at (0.3*\ix,0.6*\iy)[rotate=0,anchor=north]{{$\text{T}_\tv \mathbb{S}^2$}};
    \end{tikzpicture}
    \caption{Diagram of the Riemannian gradient. For a point $\tv$ on the manifold $\mathbb{S}^2$ and assuming a vector $\vv$ on the tangent plane $\text{T}_\tv \mathbb{S}^2$, we have ${\left\langle { {{\rm{grad}}{\cal L}} ({\tv}),{\vv}} \right\rangle }_{\tv} = {\left\langle { {{\rm{grad}} \overline {\cal L}} ({\tv}),{\vv}} \right\rangle }_{\tv} $. We can decompose ${\rm{grad}} \overline {\cal L}$ (the blue arrow) into ${\rm{grad}} \overline {\cal L}\left( {\bf{t}} \right)_\parallel$ (parallel to $\text{T}_\tv \mathbb{S}^2$) and ${\rm{grad}} \overline {\cal L}\left( {\bf{t}} \right)_\bot$ (perpendicular). Since $\vv$ lies in $\text{T}_\tv \mathbb{S}^2$, its inner product with ${\rm{grad}} \overline {\cal L}\left( {\bf{t}} \right)_\bot$ is zero, showing that ${\rm{grad}}{\cal L}\left( {\bf{t}} \right) = {\rm{grad}} \overline {\cal L}\left( {\bf{t}} \right)_\parallel$.
    }
    \label{fig2}
\end{figure}
where $\overline {\cal L} \left( {\zv} \right)$ is the smooth extension of ${\cal L} \left( {\zv} \right)$, ${{\rm{Proj}}_{\zv}}: \mathbb{R} \rightarrow \Tm_\zv {\mathbb{Q}}$ is the orthogonal projector from $\mathbb{R}$ to $\Tm_\zv {\mathbb{Q}}$, and ${\rm{grad}}\overline {\cal L} \left( {\bf{t}} \right)$ is the Euclidean gradient of likelihood function. Taking the direction vector $\tv \in \mathbb{S}^2$ as an example, the Riemannian gradient of ${\cal L} \left( {\bf{t}} \right)$ is shown in Fig. \ref{fig2}. According to the character of the orthogonal projection of manifold $\mathbb{S}^2$ \cite[(3.40)]{boumal2023introduction}, \eqref{3-5} can be written as
\begin{equation}
{\rm{grad}}{\cal L}\left( {\bf{t}} \right) = {\rm{grad}}\overline {\cal L} \left( {\bf{t}} \right) - \left[ {{{\bf{t}}^\top} \cdot {\rm{grad}}\overline {\cal L} \left( {\bf{t}} \right)} \right] \cdot {\bf{t}},
\label{3-6}
\end{equation}
For parameters in the Euclidean space, the corresponding Riemannian gradient equals the Euclidean gradient.

The ICRB of $\zv$ is defined as:
\begin{equation}
    \text{ICRB} \buildrel \Delta \over = \left[ {\Fm^{-1}_{\zv, \text{ICRB}}} \right].
\label{definition of icrb}
\end{equation}

For a Euclidean manifold, the logarithmic map simplifies to ${{\text{Log} }_\zv }(\hat \zv ) = \hat \zv - \zv$. Besides, there’s no curvature to account for. Consequently, the ICRB is equivalent to the standard CRB.


\subsection{FIM of the Channel Geometrical Parameters}
\label{3-C}
The channel geometrical parameter vector $\zv \in \mathbb{R}^{3N} \times (\mathbb{S}^2)^{2N} $ contains $N$ time delays, ${2N}$ channel gains, and ${2N}$ direction vectors. Therefore, the manifold $\mathbb{Q} = \mathbb{R}^{3N} \times (\mathbb{S}^2)^{2N} $.


\subsubsection{Riemannian metric}
Since $\mathbb{S}^2$ is embedded in $\mathbb{R}^3$, it is endowed with the inherited metric of Euclidean space on each tangent space $\text{T}_\tv \mathbb{S}^2$ as:
\begin{equation}
  {\langle \mv, \nv\rangle}_\tv = \langle \mv, \nv \rangle = \mv^\top \nv.
  \label{riemannian metric}
\end{equation}
Following the property of Riemannian metric \cite[(Example 3.57)]{boumal2023introduction}, we have
\begin{equation}
  {\left\langle {\left( {{{\bf{m}}_1}, \ldots ,{{\bf{m}}_{5N}}} \right),\left( {{{\bf{n}}_1}, \ldots ,{{\bf{n}}_{5N}}} \right)} \right\rangle _{\left( {{{\zv}_1}, \ldots ,{{\zv}_{5N}}} \right)}} = \sum\limits_{k = 1}^{5N} {\left\langle {{{\bf{m}}_k},{{\bf{n}}_k}} \right\rangle },
  \label{3-4}
\end{equation}
for $\left( {{{\bf{m}}_1}, \ldots ,{{\bf{m}}_{5N}}} \right),\left( {{{\bf{n}}_1}, \ldots ,{{\bf{n}}_{5N}}} \right) \in {\text{T}_{{\zv}_1} \mathbb{R}} \times \cdots \times {\text{T}_{{\zv}_{3N}} \mathbb{R}} \times {\text{T}_{{\zv}_{3N+1}} \mathbb{S}^2} \times \cdots {\text{T}_{{\zv}_{5N}} \mathbb{S}^2} $ and ${\left( {{{\zv}_1}, {{\zv}_2}, \ldots ,{{\zv}_{5N}}} \right)} \in \mathbb{Q} $. Therefore, we can express the calculation of the complex Riemannian metric in \eqref{5-37} as the sum of $5N$ simple Riemannian metrics.

\begin{figure}
    \centering
    \begin{tikzpicture}
    \node (image) [anchor=south west]{\includegraphics[width=0.8\linewidth]{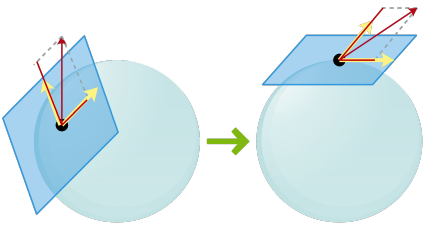}};
    \gettikzxy{(image.north east)}{\ix}{\iy};
    \node at (0.15*\ix,0.9*\iy)[rotate=0,anchor=north]{{$\red{\mv}$ }};
    \node at (0.03*\ix,0.72*\iy)[rotate=0,anchor=north]{{$\red{\mv^\top \ev_2}$ }};
    \node at (0.25*\ix,0.53*\iy)[rotate=0,anchor=north]{{$\red{\mv^\top \ev_1}$ }};
    \node at (1*\ix,1*\iy)[rotate=0,anchor=north]{{$\red{\mv}$ }};
    \node at (0.8*\ix,1*\iy)[rotate=0,anchor=north]{{$\red{\mv^\top \ev_2}$ }};
    \node at (0.83*\ix,0.72*\iy)[rotate=0,anchor=north]{{$\red{\mv^\top \ev_1}$ }};
    \node at (0.1*\ix,0.57*\iy)[rotate=0,anchor=north]{{$\yellow{\ev_2}$ }};
    \node at (0.27*\ix,0.66*\iy)[rotate=0,anchor=north]{{$\yellow{\ev_1}$ }};
    \node at (0.8*\ix,0.87*\iy)[rotate=0,anchor=north]{{$\yellow{\iv_2}$ }};
    \node at (0.95*\ix,0.8*\iy)[rotate=0,anchor=north]{{$\yellow{\iv_1}$ }};
    \node at (0.12*\ix,0.28*\iy)[rotate=0,anchor=north]{{$\text{T}_\tv \mathbb{S}^2$}};
    \node at (0.65*\ix,0.73*\iy)[rotate=0,anchor=north]{{$\text{T}_\qv \mathbb{S}^2$}};
    \node at (0.75*\ix,0.77*\iy)[rotate=0,anchor=north]{{$\qv$}};
    \node at (0.17*\ix,0.44*\iy)[rotate=0,anchor=north]{{$\tv$}};
    \node at (0.52*\ix,0.5*\iy)[rotate=0,anchor=north]{{$\Rm_{\tv}$}};
    \end{tikzpicture}
    \caption{Diagram of special orthonormal basis. $\tv \in \mathbb{S}^2$ and $\mv \in \text{T}_\xv \mathbb{S}^2 $. $\ev_1$ and $\ev_2$ (yellow arrows) form an orthonormal basis of $\text{T}_\tv \mathbb{S}^2$. ${\mv^\top \ev_1}$ and ${\mv^\top \ev_2}$ (the lengths of the red lines) represent the Riemannian metric between $\mv$ and basis. Using $\Rm_k$, we can rotate $\text{T}_\tv \mathbb{S}^2$ to the tangent space $\text{T}_\qv \mathbb{S}^2$ at the north pole $\qv$. The special orthogonal basis of $\text{T}_\qv \mathbb{S}^2$ consists of $\iv_1 = [1,0]^\top$ and $\iv_2 = [0,1]^\top$. Then, $ {\mv^\top \ev_i} = \langle \ev_i^\top  \mv , \iv_i \rangle$.
    }
    \label{fig1}
\end{figure}

The time delays and channel gain are real numbers in Euclidean space, so their corresponding orthonormal basis used in \eqref{5-37} are the orthonormal basis in Euclidean space. While the direction vector is constrained on the unit circle, the corresponding tangent space and orthonormal basis vary with the value of the direction vector. To simplify \eqref{5-37} for direction vector, we take one direction vector $\tv \in \mathbb{S}^2$ as example and introduce the rotation matrix $\Rm_{\tv} = [\ev_1, \ev_2, \tv]$, where ${\ev_1, \ev_2}$ form an orthonormal basis of $\text{T}_\tv \mathbb{S}^2$. Note that $\Rm_{\tv}$ here is not the rotation matrix of the UE; $\Rm_{\tv}$ is used to simplify the FIM calculation. Given an arbitrary $\mv$ in $\text{T}_\tv \mathbb{S}^2$, the $z$-coordinate of $\Rm_{\tv}^\top \mv = 0$.~\footnote{$\tv$ is a point on $\mathbb{S}^2$ and also represents the vector from the origin to $\tv$, making $\tv$ perpendicular to the tangent space $\text{T}_\tv \mathbb{S}^2$. Since $\mv$ is a vector in $\text{T}_\tv \mathbb{S}^2$, the $z$-coordinate of $\Rm_{\tv}^\top \mv$ is $\tv^\top \mv = 0$.} Hence, $\Rm_{\tv}^\top \mv$ lies in the tangent space of the north pole $\qv = [0,0,1]^\top$, 
which is illustrated in Fig. \ref{fig1}. 
Then, the orthonormal basis of $\text{T}_\qv \mathbb{S}^2$ can be reduced to a special orthonormal basis $[\iv_1, \iv_2]$ where $\iv_1 = [1,0]^\top$ and $\iv_2 = [0,1]^\top$. Defining a transformation matrix $\Bm_\tv = [\ev_1, \ev_2]^\top \in \mathbb{R}^{2\times3}$, we have

\begin{equation}
    {\langle \mv, \ev_i \rangle} = \langle \Bm_{\tv}  \mv , \iv_i \rangle, \text{for }\mv \in \text{T}_\tv \mathbb{S}^2.
    \label{5-38}
\end{equation}
Therefore, this operation not only simplifies the orthonormal basis but also reduces the dimension of the FIM, thereby aligning it with the degrees of freedom of the direction vector. Applying \eqref{5-38} to the Riemannian gradient ${\rm{grad}}{\cal L}\left( {\tv} \right) $ in \eqref{3-5}, the transformed Riemannian gradient is defined as
\begin{equation}
\overline {{\rm{grad}}{\cal L}} \left( {\bf{t}} \right) = \text{B}_{\tv} {\rm{grad}}{\cal L}\left( {\bf{t}} \right) \in \mathbb{R}^2,
\label{3-7-t}
\end{equation}
which represents the projection of the Riemannian gradient ${\rm{grad}}{\cal L}\left( {\bf{t}} \right)$ onto the two orthonormal basis of $\text{T}_\tv \mathbb{S}^2$.
For $\tau \in \mathbb{R}$, the transformed Riemannian gradient is equal to the Euclidean gradient.

Then, the transformed Riemannian gradient for $\zv$ is
\begin{equation}
\overline {{\rm{grad}}{\cal L}} \left( {\bf{z}} \right) = \text{B}_{\zv} {\rm{grad}}{\cal L}\left( {\bf{z}} \right) \in \mathbb{R}^{7N \times 9N},
\label{3-7}
\end{equation}
where $\Bm_{\zv} = \text{diag} \left[ {{\bf{I}}_{3N}}, \Bm_{\zv_{3N+1}} , \dots , \Bm_{\zv_{5N}} \right] \in \mathbb{R}^{7N \times 9N}$.
\subsubsection{Derivation of the FIM}
Using the transformed Riemannian gradient, the FIM of the channel geometric parameter vector \eqref{5-37} can be simplified as:
\begin{subequations}
\begin{align}
& {\left[ {{{\bf{F}}_{\zv}}} \right]_{i,j}} = \mathbb{E}\left\{ {\left\langle { \Bm_{\zv}{{\rm{grad}}{\cal L}} ({\zv}),{\iv_i}}\right\rangle_{\zv} \cdot  {{\left\langle { \Bm_{\zv}{{\rm{grad}}{\cal L}} ({\zv}),{\iv_j}} \right\rangle }_{\zv}^\top}} \right\} \label{3-8bb}\\ 
 &= \mathbb{E}\left\{ {\bf{i}}_i^\top (\overline {{\mathop{\rm grad} } {\cal L}}(\zv)) (\overline {{\mathop{\rm grad} } {\cal L}}(\zv))^\top {\bf{i}}_j \right\} \label{3-8c}\\
 &= {\bf{i}}_i^{\top} \mathbb{E}\left\{ {(\overline {{\mathop{\rm grad} } {\cal L}} (\zv)){{(\overline {{\mathop{\rm grad} } {\cal L}} (\zv))}^\top}} \right\}{{\bf{i}}_j} \label{3-8d}\\
 &= {\left[ {\mathbb{E}\left\{ {(\overline {{\mathop{\rm grad} } {\cal L}} (\zv)){{(\overline {{\mathop{\rm grad} } {\cal L}} (\zv))}^{\top} }} \right\}} \right]_{i,j}}  \in \mathbb{R}^{7N \times 7N }  \label{3-8e},
\end{align}
\end{subequations}
where ${\bf{i}}_i$ represents a column vector with the $i$-th element equal to 1 and all other elements equal to 0.
Equation \eqref{3-8bb} holds due to \eqref{5-38}, \eqref{3-8c} holds due to the character of Riemannian metric, and \eqref{3-8d} and \eqref{3-8e} holds due to the property of orthonormal basis. $\{ \cdot \}_{i,j}$ represents the element in row $i$ and column $j$.

For a direction vector $\tv$ of the received signal constrained on $\mathbb{S}^2$, the orthonormal basis $\ev_1$ and $\ev_2$ are all orthogonal to $\text{T}_\tv \mathbb{S}^2$. We can derive the transformed Riemannian gradient of $\tv$ using \eqref{3-6} and \eqref{3-7-t} as:
\begin{equation}
\overline {{\rm{grad}}{\cal L}} \left( {\bf{t}} \right) = [ \ev_1^\top \cdot {\rm{grad}}{\overline {\cal L}}\left( {\bf{t}} \right) , \ev_2^\top \cdot {\rm{grad}}{\overline {\cal L}}\left( {\bf{t}} \right) ].
\label{5-42}
\end{equation}
Substituting the above equation into \eqref{3-8e}, we can obtain:
\begin{equation}
	{\Fm_{\zv, \text{ICRB}}} = \Bm_\zv {\Fm_{\zv, \text{CRB}}} \Bm_\zv^\top,
\label{5-43}
\end{equation}
where ${\Fm_{\zv, \text{CRB}}} \in \mathbb{R}^{9N \times 9N } $ represents the unconstrained FIM.
Therefore, the FIM associated with the ICRB coincides with the conventional FIM for states in Euclidean space. For constrained direction vectors, however, this ICRB-associated FIM reduces to the projection of the unconstrained FIM onto the tangent space of the constraint manifold at the specified direction vector. The unconstrained FIM for the direction vectors can be derived from the signal model, with the detailed derivation provided in the appendix \ref{C}.

We only need to consider the channel parameters that can assist localization. The nuisance parameters, i,e, channel gains, can be removed after equivalent FIM (EFIM) calculation \cite[(72), (73)]{chen2022tutorial}. Therefore, in the next subsection, $\zv$ only includes time delays and direction vectors. ${\mathbb{Q}}$ is adjusted as $\mathbb{R}^{N} \times (\mathbb{S}^2)^{2N} $ and $\Bm_{\zv}$ is adjusted as $\text{diag} \left[ {{\bf{I}}_{N}}, \Bm_{\zv_{3N+1}} , \dots , \Bm_{\zv_{5N}} \right] \in \mathbb{R}^{5N \times 7N}$.

\subsection{ICRB of The State}
Given the FIM of the channel geometry parameter $\zv$, we can compute the FIM of UE state $\xv$, accounting for the constraints on the UE orientation. Note that the logarithmic map in the covariance matrix corresponding to the ICRB of the rotation matrix measures residual errors represented in the tangent space, which means that the ICRB bounds the residual error of the corresponding Lie algebra of the rotation matrix. Therefore, in the computation of ICRB, we need to map the rotation matrix onto its corresponding tangent space, converting it into a Lie algebra.

Denoting the mapping from $\xv$ to $\zv$ by $\xv \to \zv(\xv)$: $\mathbb{R}^3 \times \text{SO}(3) \to \mathbb{Q}$ and following the chain rule, we have the FIM of UE state vector x as:
\begin{subequations}
\begin{align}
&{\left[ {{{\bf{F}}_\xv}} \right]_{i,j}} = \mathbb{E}\left\{ {{\rm{D}}({\cal L} \circ \zv)(\xv)\left[ {{{\bf{e}}_i}} \right] \cdot {\rm{D}}({\cal L} \circ \zv)(\xv)\left[ {{{\bf{e}}_j}} \right]} \right\} \label{3-9a}\\
& = \mathbb{E}\left\{ {{\rm{D}}{\cal L}(\zv)\left\{ {{\rm{D}}\zv(\xv)\left[ {{{\bf{e}}_i}} \right]} \right\} \cdot {\rm{D}}{\cal L}(\zv)\left\{ {{\mathop{\rm D} } \zv(\xv)\left[ {{{\bf{e}}_j}} \right]} \right\} } \right\} \label{3-9b}\\
& = \mathbb{E}\left\{ {{{\left( {{\rm{D}}\zv(\xv)\left[ {{{\bf{e}}_i}} \right]} \right)}^{\top} }({\mathop{\rm grad} } {\cal L}(\zv)){{({\mathop{\rm grad} } {\cal L}(\zv))}^{\top} }\left( {{\rm{D}}\zv(\xv)\left[ {{{\bf{e}}_j}} \right]} \right)} \right\} \label{3-9c}\\
 & = \mathbb{E}\left\{ {{{\left( {\overline {{\mathop{\rm D}} \zv(\xv)} \left[ {{{\bf{e}}_i}} \right]} \right)}^{\top} }(\overline {{\mathop{\rm grad} } {\cal L}} (\zv)){{(\overline {{\mathop{\rm grad} } {\cal L}} (\zv))}^{\top} }\left( {\overline {{\mathop{\rm D} } \zv(\xv)} \left[ {{{\bf{e}}_j}} \right]} \right)} \right\} \label{3-9d}\\
 & = \iv_i^{\top} {\bf{T}}_\zv^{\top} \mathbb{E}\left\{ {(\overline {{\mathop{\rm grad} } {\cal L}} (\zv)){{(\overline {{\mathop{\rm grad} } {\cal L}} (\zv))}^{\top} }} \right\}{{\bf{T}}_\zv}{\iv_j} \label{3-15a}\\
 & = \iv_i^{\top} {\bf{T}}_\zv^{\top} {{\text{FIM}}_{\zv}}{{\bf{T}}_\zv}{\iv_j}. \label{3-15b}
\end{align}
\label{3-9}
\end{subequations}
where \eqref{3-9b} holds due to the chain rule (3.29) in \cite{boumal2023introduction}, and the differential in \eqref{3-9c} is calculated according to \eqref{grad} and \eqref{riemannian metric}. $\overline {{\rm{D}} \zv(\xv)} \left[ {{{\bf{e}}_i}} \right]$ in \eqref{3-9d} is the transformed differential defined similarly as \eqref{3-7}, given by
\begin{equation}
 \overline {{\rm{D}} \zv(\xv)} \left[ {{{\bf{e}}_i}} \right] = 
\Bm_{\zv}  {\rm{D}} \zv(\xv)\left[ {{{\bf{e}}_i}} \right],
\label{3-14}
\end{equation}
which is equivalent to rotating ${{\rm{D}} \zv(\xv)} \left[ {{{\bf{e}}_i}} \right]$ to the tangent space at the north pole. Then \eqref{3-9d} holds because $\Bm_{\zv}^\top \Bm_{\zv} = \bf{I}$, as $\Bm_{\zv}$ is a block diagonal matrix composed of multiple orthogonal basis matrices. The operator ${{\bf{T}}_\zv}$ in \eqref{3-15a} is introduced such that it satisfies ${{\bf{T}}_\zv}{\iv_i} = \overline {{\rm{D}} \zv(\xv)} \left[ {{{\bf{e}}_i}} \right]$. Therefore, we have ${{\bf{F}}_\xv} = {\bf{T}}_\zv^{\top} {{\text{FIM}}_{\zv}}{{\bf{T}}_\zv}$, where ${\bf{F}}_\xv$ is the FIM of UE state vector $\xv$.


${{\rm{D}} \zv(\xv)} \left[ {{{\bf{e}}_i}} \right]$ represents the Riemannian differential of $\zv(\xv)$ along the orthonormal basis ${\bf{e}}_i$ of tangent space. Because $\mathbb{R}^3 \times {\text{SO}(3)} $ and $ \mathbb{Q}$ are submanifolds of Euclidean space, the map $\zv(\xv)$: $\mathbb{R}^3 \times {\text{SO}(3)} \to \mathbb{Q}$ admits a smooth extension: $\overline \zv({\overline{\xv}})$: $\mathbb{R}^9 \to \mathbb{R}$, where $\mathbb{R}^9$ is a neighborhood of $\mathbb{R}^3 \times {\text{SO}(3)} $ in Euclidean space. Based on the smooth extension property of the Riemannian differential \cite[(3.28), (5.8)]{boumal2023introduction}, ${ {\rm{D}} \zv({{\xv}})} \left[ {{{\bf{e}}_i}} \right] = {{\rm{D}} {\overline{\zv}}({\overline{\xv}})} \left[ {{{\bf{e}}_i}} \right]$. We represent the gradient of ${\overline{\zv}}$ respect to ${\overline{\xv}}$ as ${\nabla _{\overline{\xv}}} \overline \zv$, i.e., ${\nabla _{\overline{\xv}}} \overline \zv = {{\rm{grad}}} {\overline{\zv}}({\overline{\xv}})$. According to \eqref{grad}, ${{\rm{D}} {\overline{\zv}}({\overline{\xv}})} \left[ {{{\bf{e}}_i}} \right] = \left\langle { {\nabla _{\overline{\xv}}} \overline \zv,{\ev_i}}\right\rangle_{{\overline{\xv}}}$, where ${\ev_i}$ is the $i$-th orthogonal basis of tangent space $\textbf{T}_{{\xv}}[\mathbb{R}^3 \times {\text{SO}(3)} ]$. For the position in $\mathbb{R}^3$, the orthogonal bases form $\mathbf{I}_3$. For the rotation matrix $\Rm$ in the state, the orthonormal bases of $\textbf{T}_{\Rm}{\text{SO}(3)}$ can be obtained by rotating the orthonormal bases of $\textbf{T}_{\mathbf{I}_3}{\text{SO}(3)}$ \cite[(7.32)]{boumal2023introduction}. Since ${\mathbf{I}_3}$ is identity matrix, the $i$-th orthonormal basis of $\textbf{T}_{\mathbf{I}_3}{\text{SO}(3)}$ is ${\text{vec}\left( {{ \iv_i^ \wedge } } \right)}$. Then we can obtain the orthonormal bases of $\textbf{T}_{\Rm}{\text{SO}(3)}$ by ${\text{vec}\left( {{ \iv_i^ \wedge } \Rm } \right)}$. Finally, these derivations yield  $\overline {{\rm{D}} \zv({\xv})} \left[ {{{\bf{e}}_i}} \right]$ in \eqref{3-5}, with the detailed derivation of ${\nabla _{\xv}} \overline \zv$ provided in Appendix \ref{A}.
\begin{figure*}[ht]
\begin{equation}
\begin{aligned}
{{\bf{T}}_\zv} {\iv_j} = \Bm_{\zv} \cdot {\nabla _\xv} \overline \zv^\top  
\cdot \left[ {\begin{array}{*{20}{c}}
\mathbf{I}_3&{}&{}&{}\\
{}&{\text{vec}\left( {{{\left[ {\begin{array}{*{20}{c}}
1\\
0\\
0
\end{array}} \right]}^ \wedge } \Rm } \right)}&{\text{vec}\left( {{{\left[ {\begin{array}{*{20}{c}}
0\\
1\\
0
\end{array}} \right]}^ \wedge } \Rm } \right)}&{\text{vec}\left( {{{\left[ {\begin{array}{*{20}{c}}
0\\
0\\
1
\end{array}} \right]}^ \wedge }\Rm } \right)}
\end{array}} \right] {\iv_j}
\end{aligned}
\label{3-10}
\end{equation}
\hrulefill
\end{figure*}

The ICRB on $\xv$ can be written as
\begin{equation}
    \text{ICRB} \buildrel \Delta \over = {\bf{F}}_\xv^{-1}.
\label{3-16}
\end{equation}
From the ICRB, the position error bound (PEB) and rotation matrix error bound (RMEB) can be written as
\begin{equation}
\begin{aligned}
    \text{PEB} &= \sqrt{\text{tr}([\text{ICRB}]_{1:3,1:3})},\\
    \text{RMEB} &= \sqrt{\text{tr}([\text{ICRB}]_{4:6,4:6})},
\end{aligned}
\end{equation}
where RMEB bounds the Lie algebra corresponding to the rotation matrix. For position vectors belonging to Euclidean space, the PEB computed by the ICRB is the same as the PEB corresponding to the unconstrained CRB.

\section{6D Filter Design}
The ICRB provides a lower bound on estimation accuracy at a single time step, quantifying the uncertainty in 6D state estimates based on the received signals. In a tracking filter operating over multiple time steps, the uncertainty of state is updated iteratively as kinematic model are incorporated. The ICRB feeds into the filter’s measurement updating process for minimizing the overall estimation error, linking the single-step ICRB to the multi-step filtering process.

For the 6D state filter, the kinematic and observation models are highly nonlinear, especially in the case of rotational dynamics represented by a rotation matrix. 
When directly tracking the rotation matrix with existing filters, the estimated states may no longer satisfy the orthogonality of the rotation matrices.
In this section, we linearize the 6D state kinematics of moving UE using the perturbation model to predict its state. We also propose two different 6D tracking filters for state updating separately based on pose fusion and error-state Kalman filter, which maintain the properties of rotation matrices.
\subsection{State Space}
The UE state $\Tm$ includes position and rotation matrix, which is defined in \eqref{state}. We can decompose the state $\Tm$ into a noise-free nominal value ${\overline \Tm }$ and a small zero-mean perturbation $\text{exp}(\delta {\bm{\xi}} ^ \wedge )$ added to its left-hand side \cite{barfoot2024state} as follows (we omit the subscript $\text{U}$ for the notation simplicity):
\begin{equation}
\begin{aligned}
\Tm = \text{exp}(\delta {\bm{\xi}} ^ \wedge ){\overline \Tm } \approx ( \mathbf{I} + \delta \bm{\xi} ^ \wedge) {\overline \Tm }
\label{T_pertur}
\end{aligned}
\end{equation}
$\Tm$ is a concentrated Gaussian distribution centered \cite{1668246} around ${\overline \Tm }$. The mean of state that propagates over time is the nominal value ${\overline \Tm }$. The corresponding covariance matrix of ${\bm{\xi}}$ is $\Pm = E[\delta {\bm{\xi} } \delta {\bm{\xi}^\top}] \in \mathbb{R}^{6 \times 6}$. So we can denote $ \delta {\bm{\xi}} \sim \mathcal {N} \left( 0, \Pm \right)$.

\subsection{Prediction Phase}
The kinematic model of the UE state for state prediction is evidently highly nonlinear. Directly linearizing the model through Taylor expansion can result in estimated rotation matrices losing orthogonality. Thus, we adopt Lie algebra's perturbation model for the coefficient matrix of the kinematic model to realize kinematic model linearization. 

According to the kinematic model \eqref{kine}, the coefficient matrix is:
\begin{equation}
\begin{aligned}
{\Fm}_k = \left[ {\begin{array}{*{20}{c}}
{\text{exp} (\Delta _k \wv_k  ^ \wedge)}&{\Delta _k \vv_k}\\
{\bf{0}}^\top&1
\end{array}} \right],
\end{aligned}
\label{F}
\end{equation}
To describe the derivations more clearly, we use $(\cdot)^{-}$ and $(\cdot)^{+}$ to represent the prior estimate and posterior estimate, respectively. The prior estimate can be updated according to \cite[(7.222), (7.253)]{barfoot2024state}:
\begin{equation}
    { {\bf{T}} _k ^{-}} = {\Fm_k}{ {\bf{T}} _{k - 1}^{+}}.
    \label{4-7}
\end{equation}
The covariance matrix $\Pm_{k}^{-} = E[\delta {\bm{\xi}_k^{-} } \delta {\bm{\xi}_k^\top}]$ is calculated by :
\begin{equation}
    \Pm_{k}^{-} =  {\text{Ad}({\Fm_k})}  \Pm_{k-1}^{+}  {{\text{Ad}({\Fm_k})}}^\top + \Qm_{{\bf{T}},k},
\label{4-8}
\end{equation}
where ${\text{Ad}(\cdot)}$ represents the adjoint matrix introduced in \eqref{Ad(SE3)} and $\Qm_{{\bf{T}},k}$ is the covariance matrix of process noise.

\subsection{Measurement Update Phase}
We consider two methods to update the state: direct fusion and the error-state Kalman filter. Direct fusion combines measurement data directly to the predicted state, dealing with errors and uncertainties between different data sources accurately in ${\text{SO}(3)}$ where the measurement models are nonlinear. On the other hand, the error-state Kalman filter focuses on the deviation of the error state by linearizing around the error state rather than the global state, this method preserves the geometric properties of the manifold and reduce computational overhead. Notably, the fusion method degenerates into a normal Kalman filter when the state resides in Euclidean space. These two methods provide complementary approaches depending on the computational resource and the accuracy requirement.
\subsubsection{Fusion Method}
\begin{figure}
    \centering
    \begin{tikzpicture}
    \node (image) [anchor=south west]{\includegraphics[width=1\linewidth]{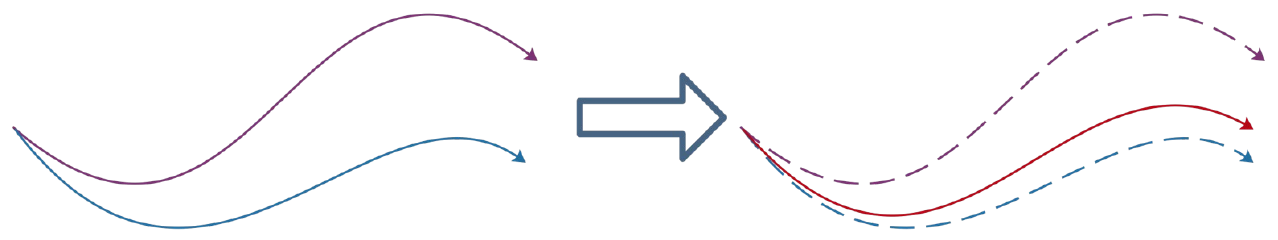}};
    \gettikzxy{(image.north east)}{\ix}{\iy};
    \node at (0.18*\ix,1*\iy)[rotate=0,anchor=north]{{$\Tm_k^{-}, \Pm_k^{-}$}};
    \node at (0.35*\ix,0.35*\iy)[rotate=0,anchor=north]{{$\hat{\Tm}_k, \hat{\Pm}_k$}};
    \node at (0.9*\ix,0.85*\iy)[rotate=0,anchor=north]{{$\Tm_k^{+}, \Pm_k^{+}$}};
    \end{tikzpicture}
    \caption{Diagram of the measurement update based on fusion. $\Tm_k^{-}$ and $\Pm_k^{-}$ represent the predicted state and its covariance, respectively, before incorporating the new measurement $\hat{\Tm}_k$ and its covariance $\Pm_k^{-}$. $\Tm_k^{+}$ represents the optimal estimate after the fusion step, reflecting adjustments based on the new measurement. } 
    \label{fig_illustration}
\end{figure}
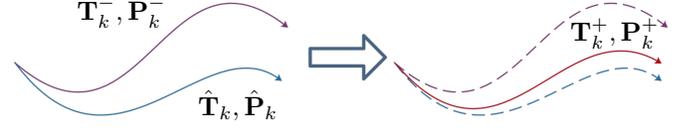

\paragraph{Main Idea}
We draw on the idea of pose fusion in robotic areas to update the state \cite{wolfe2011bayesian}, which is illustrated in Fig. \ref{fig_illustration}. At each measurement update time $k$, there exists a predicted state derived from the prediction phase and a measured state estimated using the received signal, representing two different estimates of the same state at time $k$. Both are inaccurate estimations of the true state and require fusion via specific methods to generate a more accurate estimate. Hence, the principle behind the fusion algorithm proposed in this section is to fuse the predicted and measured states to minimize the sum of the weighted errors between the updated state and both the predicted and measured states.

\paragraph{Error from The Optimal Estimate}
Assuming that the a posteriori estimate of the state is ${\Tm_{k}}^{+}$ after the update phase at time $k$, then ${\Tm_{k}}^{+}$ represents the optimal state estimate that we want to obtain at time $k$. The conventional error calculation method employed in Euclidean space is not applicable because the state belongs to the Lie group $\text{SE}(3)$. We define errors $\hv_1({{\Tm}}_{k}^{+})$ and $\hv_2({{\Tm}}_{k}^{+})  \in \mathfrak{se}(3)$, which occur between the optimal estimate ${\Tm_{k}}^{+}$ and the measurement ${\hat{\Tm}}_{k}$ and predicted state ${\Tm_{k}}^{-}$, separately:
\begin{equation}
    \hv_1({{\Tm}}_{k}^{+})=\text{log}\left[{\hat{\Tm}}_{k} ({\Tm_{k}}^{+})^{-1}\right]^\vee,
    \hv_2({{\Tm}}_{k}^{+})=\text{log}\left[{\Tm_{k}}^{-} ({\Tm_{k}}^{+})^{-1}\right]^\vee,
\label{4-13}
\end{equation}


We first give an initial guess ${\Tm_{\text{in}}}$ of the optimal estimate, then use the Gauss-Newton iteration method to continuously approach the optimal state. To maintain the orthogonality of the rotation matrix during the iteration process and to make the state always belong to $\text{SE}(3)$, the state is processed in each iteration using the perturbation method by a small amount $\boldsymbol{\epsilon}$ as
\begin{equation}
    {\Tm_{\text{op}}} = \text{exp}\left( \boldsymbol{\epsilon}^\wedge \right) {\Tm_{\text{in}}},
\label{4-14}
\end{equation}
where ${\Tm_{\text{op}}}$ represents the best estimate obtained in the current iteration. The initial values are iteratively refined by solving \eqref{4-14} several times until converging to the optimal value.

Substituting \eqref{4-14} into the definition of the error \eqref{4-13}, we can get (taking $\hv_1({{\Tm}}_{k}^{+})$ as an example):
\begin{subequations}
\begin{align}
    \hv_1({\Tm_{{\text{op}}}})&=\text{log} \left\{ {\hat{\Tm}}_{k} \left[ \text{exp}\left( \boldsymbol{\epsilon}^\wedge \right) ({\Tm_{\text{\text{in}}}}) \right]^{-1} \right\} ^\vee \label{4-15a}\\
    &=\text{log}\left[{\hat{\Tm}}_{k} ({\Tm_{\text{in}}})^{-1} \text{exp}\left( -\boldsymbol{\epsilon}^\wedge \right) \right]^\vee \label{4-15b} \\
    &= \text{log} \left\{ \text{exp} \left[ \hv_1({{\Tm}}_{\text{in}})^\wedge  \right] \text{exp} \left( -{\boldsymbol{\epsilon}}^\wedge \right) \right\}^\vee \label{4-15c}\\
    &\approx \hv_1({{\Tm}}_{\text{in}}) - \Am_1 {\boldsymbol{\epsilon}},
    \label{4-15d}
\end{align}
\label{4-15} 
\end{subequations}
where $\Am_1 = {\mathcal{J}} \left[ - \hv_1({{\Tm}}_{\text{in}}) \right] ^{-1}$. Equation \eqref{4-15c} holds based on the property of the Lie algebra $[\text{exp}\left( \boldsymbol{\epsilon}^\wedge \right)]^{-1} = \text{exp}\left( -\boldsymbol{\epsilon}^\wedge \right)$. In the derivation of \eqref{4-15d}, we use the BCH formula \eqref{3-1-7}. Since the error $\hv_1({\Tm_{\text{in}}})$ is small, it converges quickly.

\paragraph{Covariance Matrix Transformation}
The covariance matrix of $\Tm$ actually represents the error of the Lie algebra element $\gv = [{\bm{{\rho}}}^\top, {\rv}^\top ]^\top $ corresponding to $\Tm$. The covariance matrix $\hat{\Pm}$ of the measurement, given by the ICRB, describes the covariance of $\cv = [{\rv}^\top, \pv^\top ]^\top$, only with the rotation matrix component based on errors in the Lie algebra. Since $\cv$ and $\gv$ do not represent identical parameters, the covariance matrix of $\cv$ derived from the ICRB cannot be directly fused with the covariance matrix of the state $\gv$. Therefore, the covariance matrix $\hat{\Pm}$ of the measurement $\cv$ must be transformed to correspond to the format of $\Tm$.


In the covariance transformation, the subscript $k$ is omitted for simplicity in symbol representation. 
The transformed covariance matrix ${\boldsymbol{\Sigma}}_{\text{meas}}$ for measurement can be obtained from $\hat{\Pm}$ by the chain rule as \cite[(67)]{chen2022tutorial}:
\begin{equation}
    \begin{aligned}
        {\boldsymbol{\Sigma}}_{\text{meas}} = \left[ (\frac{\partial \cv}{\partial \gv}) \hat{\Pm}^{-1} (\frac{\partial \cv}{\partial \gv})^\top  \right]^{-1} , 
    \end{aligned}
    \label{SigmatoP}
\end{equation}
where ${\partial \cv} / {\partial {\gv}}$ is the Jacobian matrix that is calculated as:
\begin{equation}
    \begin{aligned}
       \frac{\partial \cv}{\partial \gv} = \left[ {\begin{array}{*{20}{c}}
\displaystyle \frac{\partial {\rv}}{\partial {\bm{{\rho}}}}  &  \displaystyle \frac{\partial {\rv}}{\partial {\rv}} \\
\displaystyle \frac{\partial \pv}{\partial {\bm{{\rho}}}}  &  \displaystyle \frac{\partial \pv}{\partial {\rv}} 
\end{array}} \right]
= \left[ {\begin{array}{*{20}{c}}
{\textbf{0}}_{3 \times 3}  &  {\textbf{1}}_{3 \times 3} \\
\Jm  &  \displaystyle \frac{\partial \pv}{\partial {\rv}} 
\end{array}} \right]
    \end{aligned}
    \label{JacobianMatrix}
\end{equation}
where $\Jm$ is defined in \eqref{3-1-5}. ${\partial \pv} / {\partial {\rv}}$ is given in the Appendix \ref{B}. Then, the obtained covariance matrix ${\boldsymbol{\Sigma}}_{\text{meas}}$ and the state's covariance matrix now represent errors for the same type of parameter. 
For consistency in notation, we denote the covariance matrix obtained in the prediction stage by ${\boldsymbol{\Sigma}}_{\text{pred}}$, i.e. ${\boldsymbol{\Sigma}}_{\text{pred}} =  \Pm_{k}^{-}$.

\paragraph{Cost Function}
We define the cost function as the sum of the weighted estimation errors, i.e \cite{mahalanobis2018generalized}:
\begin{equation}
\begin{aligned}
    C_k & = \displaystyle  \sum \limits_{m \in \{ {\text{meas, pred}} \} }  \hv_m({{\Tm}}_{k}^{+})^\top \boldsymbol{\Sigma}_m^{-1} \hv_m({{\Tm}}_{k}^{+})\\
    & \approx \displaystyle  \sum \limits_{m \in \{ {\text{meas, pred}} \} } \left[ \hv_m({{\Tm}}_{\text{in}}) - \Am_m {\boldsymbol{\epsilon}} \right]^\top \boldsymbol{\Sigma}_m^{-1}  \left[ \hv_m({{\Tm}}_{\text{in}}) - \Am_m {\boldsymbol{\epsilon}} \right],
\label{4-16}   
\end{aligned}
\end{equation}
where $\sum \limits_{m \in \{ {\text{meas, pred}} \} }$ means the sum of all measurement parts and predicted part. We simplify it as $\sum \limits_{m }$ below.
We need to find a ${\boldsymbol{\epsilon}}$ to minimize the cost function. Taking the derivative with respect to ${\boldsymbol{\epsilon}}$ and setting to zero, we can obtain the optimal ${\boldsymbol{\epsilon}}^{*}$ at each iteration as
\begin{equation}
\begin{aligned}
    {\boldsymbol{\epsilon}}^{*} = \left( \sum \limits_{m }\Am_m^{\top}  \boldsymbol{\Sigma}_m^{-1}   \Am_m \right)^{-1}
     \left[ \sum \limits_{m } \Am_m^{\top} \boldsymbol{\Sigma}_m^{-1} \hv_m({{\Tm}}_{\text{in}})  \right]
\label{4-17}   
\end{aligned}
\end{equation}
Then we update the state with the calculated ${\boldsymbol{\epsilon}}^{*}$ and use the updated state as the initial value in the next iteration, which is expressed as
\begin{equation}
\begin{aligned} 
    {\Tm_{\text{in}}} \leftarrow \text{exp}\left( {{\boldsymbol{\epsilon}}^{*}}^\wedge \right) {\Tm_{\text{in}}}
\label{4-18}   
\end{aligned}
\end{equation}

We set the threshold value of ${\boldsymbol{\epsilon}}$ for terminating the iterations, and ${\boldsymbol{\epsilon}}$ converges when it is less than the threshold. The initial value of this iteration is the optimal state, i.e., $\Tm_k^{+} = \Tm_{\text{in}}$. The corresponding covariance matrix is ${\Pm_{k}^{+}}= ( \sum \limits_{{m }} \Am_m^{\top}  \boldsymbol{\Sigma}_m^{-1}   \Am_m )^{-1} $. 

We take the tracking process of time $k$ as an example. The pseudo-codes for tracking process based on fusion can be found in Algorithm 1. 

\begin{algorithm}[t]
\caption{Fusion-based 6D Tracking} 
\hspace*{0.02in} {\bf Input:} 
\small
prior state $\Tm_{k-1}^{+}$ and covariance matrix $\Pm_{k-1}^{+}$ at time $k-1$, state measurement ${\hat{\Tm}}_{k}$ and covariance matrix ${\hat \Pm}_{k}$, iteration threshold value ${\boldsymbol{\epsilon}}_{th}$;\\
\hspace*{0.02in} {\bf Output:}
updated state $\Tm_k^{+}$ and covariance matrix $\Pm_{k-1}^{+}$;
\begin{algorithmic}[1]
\State \textbf{--- Prediction Phase ---} 
\State Estimate rotation matrix ${ {\bf{T}} _k ^{-}}$ and its covariance matrix $\Pm_{k}^{-} $ using \eqref{4-7} and \eqref{4-8}.

\State \textbf{--- Measurement Update Phase ---} 
\State Given an initial guess ${\Tm_{\text{in}}}$.
    \State Calculate the transformed covariance matrices ${\boldsymbol{\Sigma}}_{\text{mear},k}$ using \eqref{3-1-5}, \eqref{SigmatoP}, and \eqref{JacobianMatrix}, and ${\boldsymbol{\Sigma}}_{\text{pred},k}= \Pm_{k}^{-} $.
\Repeat
    \State Calculate the errors $\hv_1({{\Tm}}_{\text{in}}) $ and $\hv_2({{\Tm}}_{\text{in}}) $ using \eqref{4-13}.
    \State Calculate the Jacobian matrices $\Am_1 = {\mathcal{J}} \left[ - \hv_1({{\Tm}}_{\text{in}}) \right] ^{-1}$ and $\Am_2 = {\mathcal{J}} \left[ - \hv_2({{\Tm}}_{\text{in}}) \right] ^{-1}$ using \eqref{3-1-6}.
    \State Calculate optimal ${\boldsymbol{\epsilon}}^{*}$ using \eqref{4-17}.
    \State Update state $\Tm_{\text{in}}$ as the initial value in the next iteration using \eqref{4-18}.
\Until{${\boldsymbol{\epsilon}}^{*} < {\boldsymbol{\epsilon}}_{th}$ }
\State $\Tm_k^{+} = \Tm_{\text{in}}$, ${\Pm_{k}^{+}} = \left( \sum \limits_{m = \text{meas, pred}} \Am_m^{\top}  \boldsymbol{\Sigma}_m^{-1}   \Am_m \right)^{-1}$.
\State \Return $\Tm_k^{+}$, ${\Pm_{k}^{+}}$
\end{algorithmic}
\normalsize
\end{algorithm}

\subsubsection{Error-state Kalman Filter}
The fusion-based updating algorithm requires multiple iterations to achieve the optimal state estimate at each time, leading to high computational complexity. To address this, we use a low-complexity update algorithm based on the error-state Kalman filter \cite{madyastha2011extended} in this subsection.
\paragraph{Main Idea}
We can estimate the position and rotation matrix of UE at each snapshot by the transmitted signal, so we consider the measurement model at time $k$ as:
\begin{equation}
    \hat{\Tm}_k = \text{exp} ({\nv_k} ^\wedge) \cdot \Tm_k
\label{4-C-2-1}
\end{equation}
where $\hat{\Tm}_k$ is the measurement and ${\nv_k} \sim N(0,  {\boldsymbol{\Sigma}}_{\text{meas}}) $ represents the measurement noise.
Although the measurement model is nonlinear, the Lie algebraic error can be updated approximately linearly by means of the BCH formula. Here, we adapt the error-state Kalman filter for ${\text{SO}(3)}$ in the measurement update phase.
\paragraph{Error State and Its Measurement Model}
We separate the true state of $\Tm$ as a noise-free nominal value $\overline \Tm$ and an error state $\delta {\bm{\xi}}$, as shown in \eqref{T_pertur}.
After the prediction phase, we get the predicted nominal state ${\overline \Tm}_k^{-} = \Tm_k^{-}$. Then, the innovation term, i.e., the difference between measurement $\hat{\Tm}_k$ and prediction ${\overline \Tm}_k^{-}$ projected into the measurement space by the measurement model \cite{ribeiro2004kalman}, for ${\text{SO}(3)}$ can be defined as:
\begin{equation}
\begin{aligned}
  &  {\bm{\gamma}}_k  = \left[ \text{log} \left( \hat{\Tm}_k \cdot {{\overline \Tm}_k^{-}}^{-1} \right)  \right]^\vee \\
           & \overset{\eqref{4-C-2-1}, \eqref{T_pertur}} {=} \left\{ \text{log} \left\{ \text{exp} ({\nv_k} ^\wedge) \Tm_k \left[ { \text{exp} (-{\delta {\bm{\xi}}}_k^\wedge) {\Tm}_k  } \right]^{-1}   \right\}  \right\}^\vee \\
           &  \overset{\eqref{3-1-7}}{\approx} \nv_k  + {\boldsymbol{\mathcal{J}}}_l(\nv_k ) {\delta {\bm{\xi}}}_k
            \ = \ {\delta {\bm{\xi}}}_k + \nv_k ,
\label{4-C-2-2}
\end{aligned}
\end{equation}
where the penultimate approximate equation holds based on the fact that $ {\boldsymbol{\mathcal{J}}}_l(\nv_k )$ is approximately equal to $\bf{I}$ when $\nv_k$ small. This is a linear measurement model of $ {\delta {\bm{\xi}}}_k$.
\paragraph{Kalman Filter for Error State}
Then the classical Kalman filter can be applied to update the error state as follows \cite{madyastha2011extended}:
\begin{equation}
\begin{aligned}
    \Km_k = {\boldsymbol{\Sigma}}_{\text{pred},k}^{-} \left( {\boldsymbol{\Sigma}}_{\text{pred},k}^{-} + {\boldsymbol{\Sigma}}_{\text{meas},k} \right)^{-1},
\label{4-C-2-5(1)}
\end{aligned}
\end{equation}
\begin{equation}
\begin{aligned}
    \hat {\delta {\bm{\xi}}_k^{+}} = \Km_k \left\{ \text{log} \left[ \hat{\Tm}_k ({\Tm}_k^{-})^{-1} \right]  \right\}^\vee ,
\label{4-C-2-5(2)}
\end{aligned}
\end{equation}
\begin{equation}
\begin{aligned}
    \boldsymbol{\Sigma}_{\bm{\varepsilon} , k} = ({\bf{I}} - \Km_k) {\boldsymbol{\Sigma}}_{\text{pred},k}^{-},
\label{4-C-2-5(3)}
\end{aligned}
\end{equation}
where $\hat {\delta {\bm{\xi}}_k^{+}}$ is the posterior estimation of the error state.
\paragraph{Update the Original State}
The posterior distribution of the error state is not a zero-mean distribution according to \eqref{4-C-2-5(2)}, which does not align with the definition of the error state. Therefore, the error state needs to be corrected. In addition, the non-zero mean of the posterior error state prevents us from directly deriving the covariance matrix of the original ${\text{SO}(3)}$ state. To facilitate this calculation, we need to transform the expression of the true state. 

The random variable of the state can be represented by the prior estimate or posterior estimate, that is:
\begin{equation}
    \begin{aligned}
        \Tm = \text{exp}({\delta {\bm{\xi}}_k^{+}}^\wedge ) \overline \Tm_k^{-} = \text{exp}({\delta {\bm{\xi}}_k^{c}} ^\wedge ) \overline \Tm_k^{+} ,
    \end{aligned}
\label{4-C-2-6}
\end{equation}
where $\delta {\bm{\xi}}_k^{c}$ represents the corrected error state.
We need to implement some transformation to find an expression of the distribution of ${\delta {\bm{\xi}}_k^{c}}$ with zero means, then we can obtain the posterior estimate of $\Tm_k$. So we rewrite the middle part of \eqref{4-C-2-6} as :
\begin{equation}
    \begin{aligned}
        \text{exp}({\delta {\bm{\xi}}_k^{+}} ^\wedge ) \overline \Tm_k^{-} = \text{exp} \left[ ({\hat {\delta {\bm{\xi}}_k^{+}}} + \sv_k ) ^\wedge \right] \overline \Tm_k^{-} ,
    \end{aligned}
\label{4-C-2-7}
\end{equation}
where $\sv_k \sim N( 0 , \boldsymbol{\Sigma}_{\bm{\varepsilon} , t})$. According to the deformation of the BCH formula, we can get
\begin{equation}
    \begin{aligned}
        \text{exp}(\delta {\bm{\xi}}_k ^\wedge ) \overline \Tm_k^{-} \approx \text{exp} \left\{ [{\boldsymbol{\mathcal{J}}}_l (\hat {\delta {\bm{\xi}}_k^{+}}) \cdot \sv_k]^\wedge \right\} \text{exp} (\hat {\delta {\bm{\xi}}_k^{+}}^\wedge) \overline \Tm_k^{-} ,
    \end{aligned}
\label{4-C-2-8}
\end{equation}
Comparing with \eqref{4-C-2-6}, we can obtain the posterior estimate of the original state as;
\begin{equation}
\begin{aligned}
    \Tm_k^{+} = \text{exp} ( \hat {\delta {\bm{\xi}}_k^{+}} ^\wedge) \overline \Tm_k^{-} 
\label{4-C-2-9}
\end{aligned}
\end{equation}
Meanwhile, the corresponding error state is:
\begin{equation}
\begin{aligned}
    {\delta {\bm{\xi}}_k^{c}} = {\boldsymbol{\mathcal{J}}}_l (\hat {\delta {\bm{\xi}}_k^{+}}) \sv_k
\label{4-C-2-10}
\end{aligned}
\end{equation}
Hence, the covariance matrix of the Lie algebra of the original state is:
\begin{equation}
\begin{aligned}
    {\Pm_{k}^{+}} & = \Em[{\delta {\bm{\xi}}_k^{c}} {\delta {\bm{\xi}}_k^{c}}^\top ]  = \Em[{\boldsymbol{\mathcal{J}}}_l ({\hat {\delta {\bm{\xi}}_k^{+}}}) \sv_k \sv_k^\top  {\boldsymbol{\mathcal{J}}}_l ({\hat {\delta {\bm{\xi}}_k^{+}}})^\top ] \\
     & = {\boldsymbol{\mathcal{J}}}_l ({\hat {\delta {\bm{\xi}}_k^{+}}}) \boldsymbol{\Sigma}_{\bm{\varepsilon} , t} {\boldsymbol{\mathcal{J}}}_l ({\hat {\delta {\bm{\xi}}_k^{+}}})^\top
\label{4-C-2-11}
\end{aligned}
\end{equation}

We take the tracking process of epoch $k$ as an example too. The pseudo-codes for tracking process based on error-state EKF can be found in Algorithm 2.

\begin{algorithm}[t]
\caption{Error-State Kalman Filter-based 6D Tracking} 
\hspace*{0.02in} {\bf Input:} 
\small
prior state $\Tm_{k-1}^{+}$ and covariance matrix $\Pm_{k-1}^{+}$ at epoch $k-1$, state measurement ${\hat{\Tm}}_{k}$ and covariance matrix ${\hat \Pm}_{k}$;\\
\hspace*{0.02in} {\bf Output:}
updated state $\Tm_k^{+}$;
\begin{algorithmic}[1]
\State \textbf{--- Prediction Phase ---} 
\State Same as step 2 - step 4 of Algorithm 1.

\State \textbf{--- Measurement Update Phase ---} 
    \State Calculate the transformed covariance matrices $ {\boldsymbol{\Sigma}}_{\text{meas},k}$ of Lie algebra using \eqref{3-1-5}, \eqref{SigmatoP} and \eqref{JacobianMatrix} , and $ {\boldsymbol{\Sigma}}_{\text{pred},k} = \Pm_{k}^{-} $.
    \State Calculate the Kalman gain $\Km_k$ using \eqref{4-C-2-5(1)}.
    \State Calculate the posterior estimate of state $\Tm_k^{+}$ using \eqref{4-C-2-5(2)} and \eqref{4-C-2-9}.
    \State Calculate the covariance matrix ${\Pm_{k}^{+}}$ of Lie group using\eqref{3-1-6}, \eqref{4-C-2-5(2)}, \eqref{4-C-2-5(3)} and \eqref{4-C-2-11}.
\State \Return $\Tm_k^{+}$, ${\Pm_{k}^{+}}$
\end{algorithmic}
\normalsize
\end{algorithm}

\section{Simulation Results}
\subsection{Simulation Scenario}
We use the MATLAB R2020b simulation platform to simulate and analyze the ICRB and UE position estimation algorithms derived above. We consider a 3D scenario, including a BS with $N_{\text{BS}} = 64 (8 \times 8)$ antennas and a moving UE with $N_{\text{UE}} = 16 (4 \times 4)$. The parameters are set default as follows: carrier frequency $f_c = 30$ GHz, subcarrier spacing $\Delta f = 120$ kHz, bandwidth $W = 100$ MHz, average transmission power $P = 20$ dBm, number of subcarriers $K = 100$, number of transmissions $G = 20$ (analog), noise PSD $N_0 = -173.855$ dBm/Hz. The two base stations are installed at ${\pv_{\text{B1}}} = [5,0,0]^\top$ and ${\pv_{\text{B2}}} = [0, 5 ,0]^\top$, with the respective initial orientations $\Rm_{\text{B1}} = [0,15,0]^\top$ and $\Rm_{\text{B2}} = [-30,15,0]^\top$. The initial position and orientation of UE is set as ${\pv_{\text{U}}} = [-5,-5,0]^\top$ and $\Rm_{\text{U}} = [20,-30,0]^\top$. The measurements of the channel geometric parameters are estimated by the ESPRIT algorithm, which closely approaches the CRB, so that the standard CRBs for position and Euler angles are also obtained by the channel estimator. 

\subsection{Verify the ICRB}
We begin by assessing the effect of various signal parameters on the performance bounds of the single snapshot state estimation. For comparison purposes, the system is assumed to be synchronized. Employing a fully connected antenna array, each RF chain transmits different random data symbols with normalized energy. Assuming that there is no a priori information about the UE, the angle of the beam assignment is set randomly across transmissions.

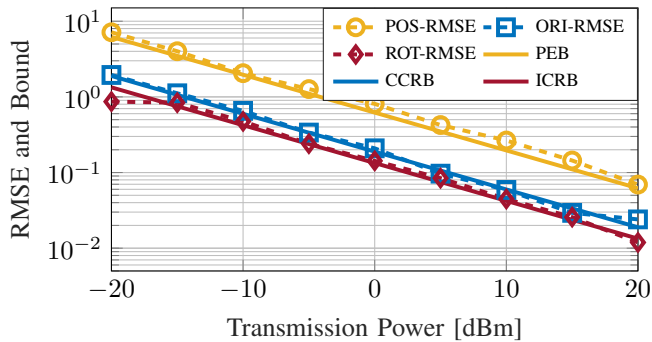
\begin{figure}
\centering
\begin{minipage}[h]{0.97\linewidth}
\centering
%
%
\definecolor{mycolor1}{rgb}{0.92941,0.69412,0.12549}%
\definecolor{mycolor2}{rgb}{0.46667,0.67451,0.18824}%
\definecolor{mycolor3}{rgb}{0.00000,0.44706,0.74118}%
\definecolor{mycolor4}{rgb}{0.63529,0.07843,0.18431}%
\begin{tikzpicture}

\begin{axis}[%
width=7cm,
height=3.5cm,
at={(0in,0in)},
scale only axis,
xmin=-20,
xmax=20,
xlabel style={font=\color{white!15!black}},
xlabel={Transmission Power [dBm]},
ymode=log,
ymin=0.005,
ymax=15,
yminorticks=true,
ylabel style={font=\color{white!15!black}},
ylabel={RMSE and Bound},
axis background/.style={fill=white},
xmajorgrids,
ymajorgrids,
yminorgrids,
legend style={at={(1,1)}, legend cell align=left, align=left, font = \scriptsize, legend columns = 2, draw=white!15!black}
]
\addplot [color=mycolor1, line width=1.5pt, dashed, mark=o, mark size=3.0pt, mark options={solid, mycolor1}]
  table[row sep=crcr]{%
-20 7.1441890257709\\
-15	4.01067231822531\\
-10	2.05766356793551\\
-5	1.26800440212765\\
0	0.80860227786939\\
5	0.424592060518767\\
10	0.264955906862496\\
15	0.144234758698204\\
20	0.069337496749239\\
};
\addlegendentry{POS-RMSE}

\addplot [color=mycolor3, line width=1.5pt, dashed, mark=square, mark size=3.0pt, mark options={solid, mycolor3}]
  table[row sep=crcr]{%
-20 1.94777696104632\\
-15	1.11565800281439\\
-10	0.653423551709440\\
-5	0.337476840114394\\
0	0.209028248532542\\
5	0.0966273412356950\\
10	0.0588826063820699\\
15	0.0293119423020202\\
20	0.0240175289914668\\
};
\addlegendentry{ORI-RMSE}

\addplot [color=mycolor4, line width=1.5pt, dashed, mark=diamond, mark size=3.0pt, mark options={solid, mycolor4}]
  table[row sep=crcr]{%
-20 0.865593993450133\\
-15	0.838269938595326\\
-10	0.474645695372391\\
-5	0.240422682693580\\
0	0.142584952510598\\
5	0.0842013371952366\\
10	0.0446454387780208\\
15	0.0257273267147295\\
20	0.01191147049982671\\
};
\addlegendentry{ROT-RMSE}

\addplot [color=mycolor1, line width=1.5pt]
  table[row sep=crcr]{%
-20 6.19955533721180\\
-15	3.48626616391876\\
-10	1.96047153458418\\
-5	1.10245416075599\\
0	0.619955533721305\\
5	0.348626616391840\\
10	0.196047153458474\\
15	0.110245416075587\\
20	0.0619955533721106\\
};
\addlegendentry{PEB}

\addplot [color=mycolor3, line width=1.5pt]
  table[row sep=crcr]{%
-20 1.88952291728387\\
-15	1.06255682128297\\
-10	0.597519610970305\\
-5	0.336009969860263\\
0	0.188952291728391\\
5	0.106255682128297\\
10	0.0597519610970297\\
15	0.0336009969860266\\
20	0.0188952291728392\\
};
\addlegendentry{CCRB}

\addplot [color=mycolor4, line width=1.5pt]
  table[row sep=crcr]{%
-20 1.336094468018812\\
-15	0.751341133725194\\
-10	0.422510168809048\\
-5	0.237594928234478\\
0	0.133609446801883\\
5	0.0751341133725204\\
10	0.0422510168809051\\
15	0.0237594928234481\\
20	0.0133609446801883\\
};
\addlegendentry{ICRB}

\end{axis}
\end{tikzpicture}%
    \vspace{-0.8cm}
\end{minipage}
\caption{Error bound and RMSE at different transmission power.}
\label{fig4}
\end{figure}

We first compare the ICRB, CCRB, and tangent space error of the rotation matrix for different transmit powers, as shown in Fig. \ref{fig4}. The root mean squared error (RMSE) of the rotation matrix is calculated differently from RMSE in Euclidean space by using the Lie algebra as follows. Firstly, calculate the error of the rotation matrix $\Rm_\text{noise}$ based on the estimate of the rotation matrix $\Rm_\text{est}$, i.e., $\Rm_\text{noise}=\Rm_\text{est} \cdot \Rm_\text{real}^\top$, where $\Rm_\text{real}$ represents the real value of rotation matrix. Secondly, the error of the rotation matrix is then converted to the corresponding Lie algebra $\rv_\text{noise}$ according to the conversion relation between Lie groups and Lie algebras. Finally, the root mean square RMSE of $\rv_\text{noise}$ for multiple measurements is obtained.

As depicted in Fig. \ref{fig4}, the error of the rotation matrix decreases with increasing transmit power. Compared to the CCRB, the ICRB can better capture the RMSE of the estimator in high SNR regions.
This further proves the superior accuracy of the distances captured by the ICRB over those by the CCRB. Moreover, as the transmit power decreases, the RMSE initially rises before stabilizing at a value slightly below 1. This phenomenon stems from the constraint on the rotation matrix, wherein the norms of all column vectors are constrained to 1. Consequently, there exists an upper bound on both the error rotation matrix and its corresponding Lie algebra.

\subsection{Performance Analysis of Tracking Algorithms}
In this section, we first validate the effectiveness of the proposed two algorithms, comparing their performance. Then we compare the proposed algorithm with the Euler angle-based tracking algorithm commonly used for orientation tracking.

\begin{figure}
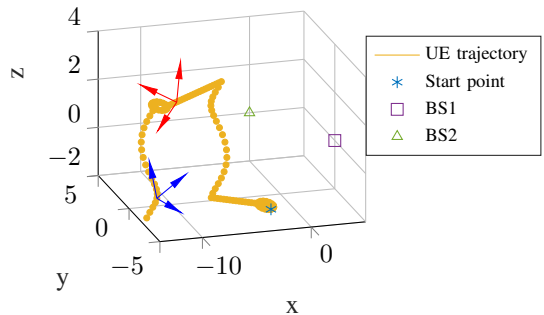

\centering
\begin{minipage}[h]{0.97\linewidth}
\centering
    \include{Figures_tikz/fig6}
    \vspace{-0.8 cm}
\end{minipage}
\caption{Simulated scenario with two fixed BSs and a moving UE. UE is moving on the track as indicated by yellow points. The red and blue coordinate systems represent the local coordinate systems induced by the user's rotation at two different points in the trajectory.}
\label{fig6}
\end{figure}
The trajectory of the UE is shown as the yellow point in Fig. \ref{fig6}, where the start point is marked by the blue star. During the movement of the UE, the sampling interval is 0.5 s and the total number of samples is 120. To validate the effectiveness of the proposed algorithm across various user motion trajectories, we segment the trajectory equally into six sections, including three types of motion: straight-line motion on the plane, approximate circular motion on the plane, and spiral ascent and descent. The translation velocities are $\vv = [0.5,0,0]^\top$, $\vv = [0,0.5,0]^\top$, $\vv = [-0.5,0,0.5]^\top$, $\vv = [0.5,0.5,0]^\top$, $\vv = [0,-0.5,0]^\top$, and $\vv = [-0.5,0,-0.5]^\top$. To ensure that the motion induced by the rotational velocity of the Euler angles matches that of the rotation matrix, we set the rotational velocity to be $\wv = [0,0,-\pi/4 ]^\top$, $\wv = [0,0,0]^\top$, $\wv = [0,0,-\pi/4 ]^\top$, $\wv = [0,0,0]^\top$, $\wv = [0,0,-\pi/4 ]^\top$, and $\wv = [0,0,-\pi/4 ]^\top$. The channel geometric parameters are obtained using the ESPRIT algorithm, while the UE's position and rotation matrix measurements are generated based on the ICRB derived from these parameters.

\subsubsection{Tracking Results}
We compare the tracking performance of the proposed algorithms with the Euler angle-based EKF algorithm \cite{Talvitie2023orientation}. We set the signal-to-noise ratio of the transmitted signal to 5 dB, i.e., the case of large measurement error. Performance is evaluated by the Monte Carlo simulation with 100 independent runs. 

\begin{figure}
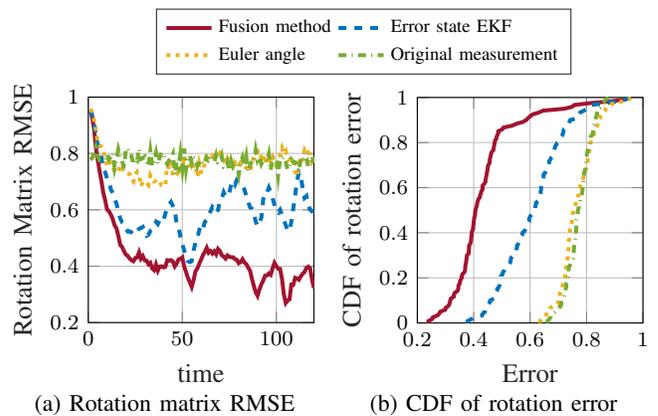

\begin{minipage}[b]{0.48\linewidth}
\centering
    \include{Figures_tikz/fig7a}
    \vspace{-1 cm}
    \centerline{\small{(a) Rotation matrix RMSE}} \medskip
\end{minipage}
\begin{minipage}[b]{0.48\linewidth}
    \include{Figures_tikz/fig7b}
    \vspace{-1 cm}
    \centerline{\small{(b) CDF of rotation error}} \medskip
\end{minipage}
\caption{Rotation matrix estimation performance of different algorithms.}
\label{fig7}
\end{figure}
The rotation matrix estimation performance of different algorithms is shown in Fig. \ref{fig7}, including rotation matrix RMSE (Fig. \ref{fig7}(a)) and CDF of rotation error (Fig. \ref{fig7}(b)). From Fig. \ref{fig7}, it is evident that the fusion method exhibits the best performance in tracking the rotation matrix. Between time 60 and time 80, there is a slight increase in the RMSE of the rotation matrix. This phenomenon can be attributed to the fact that when the value of the UE's coordinates is large, even a small error in the rotation matrix results in a larger error in the  ${{\Rm_{\text{U},k}} {\pv_{\text{U},k}}}$ term in the upper right corner of the state matrix. This term impacts the Jacobian matrix of the Lie algebra and consequently affects the accuracy of the rotation matrix estimation during the update phase. This effect is more obvious in the error-state EKF algorithm. The performance of the error-state EKF algorithm is inferior to that of the fusion algorithm due to its reliance on linearizing the measurement model, which introduces approximation errors. In contrast, the fusion algorithm avoids such approximations, resulting in lower position estimation errors. The poor estimation performance of the rotation matrix of the Euler angle-based method is due to the fact that the covariance matrix corresponding to the Euler angle into which the rotation matrix measurements are converted does not accurately characterize the covariance matrix of the rotation matrix measurements. Therefore the measurement error cannot be significantly reduced in the update phase.

\begin{figure}
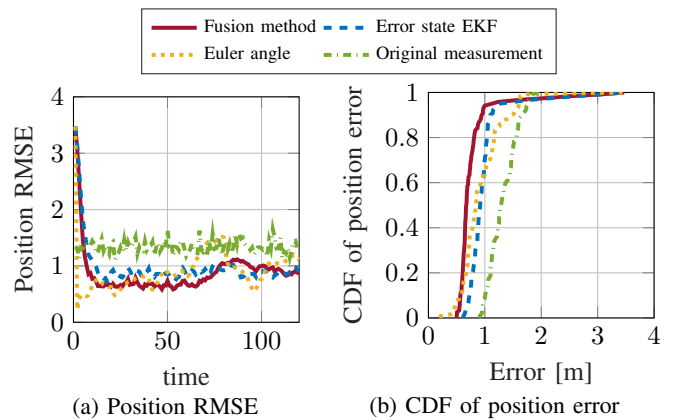

\centering
\begin{minipage}[b]{0.48\linewidth}
\centering
    \include{Figures_tikz/fig8a}
    \vspace{-1 cm}
    \centerline{\small{(a) Position RMSE}} \medskip
\end{minipage}
\begin{minipage}[b]{0.48\linewidth}
\centering
    \include{Figures_tikz/fig8b}
    \vspace{-1 cm}
    \centerline{\small{(b) CDF of position error}} \medskip
\end{minipage}
\caption{Position estimation performance of different algorithms.}
\label{fig8}
\end{figure}
In terms of position estimation, the results of position estimation are given in Fig. \ref{fig8}. The performance of the three methods is similar, as can be seen in Fig. \ref{fig8}(b), with the fusion method slightly outperforming the others. The measurement model for the position is linear, allowing the EKF to provide accurate estimates. However, the position estimation in the UE state is also influenced by the accuracy of rotation matrix estimation. Therefore, the position estimation performance of the fusion method, which accurately estimates the rotation matrix, is superior. In Fig. \ref{fig8}(a), the position estimation error of the Euler angle-based method exhibits fluctuations between time 60 and 80, which are also attributed to the larger values of position coordinates.

\section{Conclusion}
ISAC systems enable precise estimation of both position and orientation in dynamic environments using communication signals. In this work, we formulated and solved the 6D tracking problem using the ICRB. For channel geometric parameters and rotation matrices constrained on a Riemannian manifold, we derive the ICRB using Lie algebra theory, which effectively describes the covariance of rotation matrix error in the latent space. To address the nonlinear 6D state kinematics, , we linearize the motion model using a perturbation approach. To handle the orthogonality constraints and multiplicative noise of rotation matrices, we integrate the derived ICRB into two 6D tracking filters for state measurement updates: one based on the fusion method and the other on the error-state Kalman filter. Simulation results show that the derived ICRB accurately characterizes the performance of the maximum likelihood estimator, and both filters outperform the commonly used Euler angle-based tracking algorithms. Although the fusion-based filter has a longer computation time than the error-state Kalman filter-based tracking algorithm, it significantly reduces estimation errors, particularly in the rotation matrix estimation.


\begin{appendices}
\section{Derivation of the FIM of direction vector} \label{C}
The direction vector is related to the signal's angle of arrival or departure. The FIM of the angle ${\text{FIM}}_{\mathbf{\gamma }} $ can be computed from the signal model, where ${\mathbf{\gamma }}$ represents the angle vector associated with the direction vector. So the FIM of the unconstrained direction vector can be further derived according to the chain rule of the derivation. Taking one direction vector as an example, the CRB of this direction vector can be derived as:
\begin{equation}
	{\text{FIM}}_{\text{CRB,}\tv}= \Jm_{\tv}^\top {\text{FIM}}_{\mathbf{\gamma }}  
    \Jm_{\tv} \in \mathbb{R}^{3 \times 3}.
\label{5-44}
\end{equation}
where $ \Jm_{\tv}$ denotes the Jacobian matrix of the angle with respect to the direction vector, which can be written as:
\begin{equation}
{{\bf{J}}_{\bf{t}}} \buildrel \Delta \over = \displaystyle \frac{{\partial {\mathbf{\gamma }}}}{{\partial {\bf{t}}}} = \left[ {\begin{array}{*{20}{c}}
{\displaystyle \frac{{ - {{\bf{t}}_2}}}{{{\bf{t}}_1^2 + {\bf{t}}_2^2}}}&{\displaystyle \frac{{{{\bf{t}}_1}}}{{{\bf{t}}_1^2 + {\bf{t}}_2^2}}}&0\\
0&0&{{{(1 - {\bf{t}}_3^2)}^{ -  \frac{1}{2}}}}
\end{array}} \right]
\label{5-45}
\end{equation}
where $\tv_i$ denotes the $i$-th element of the vector. 

\section{Derivation of ${\nabla _\xv} \overline \zv$} \label{A}
The derivative with respect to $\pv$ is independent of that with respect to $\text{vec}(\Rm)$, so we separate ${\nabla _\xv} \overline \zv$ into two parts. For rotation matrix, the derivative of the direction vector with respect to the vectorized rotation matrix is obtained using \eqref{9}:
\begin{equation}
    {\nabla _{\text{vec}(\Rm)}} \overline \tv_i = 
    \begin{bmatrix}
    \Rm_\text{U} \tv_i & \mathbf{0}_3 & \mathbf{0}_3\\
    \mathbf{0}_3 & \Rm_\text{U} \tv_i & \mathbf{0}_3\\
    \mathbf{0}_3 & \mathbf{0}_3 & \Rm_\text{U} \tv_i\\
    \end{bmatrix}^\top, i\in \{1,...,N\}
\label{3-11}
\end{equation}
For position, the derivative of the time delay with respect to the position ${\nabla _\pv} \overline \tau_i, i\in \{1,...,2N\}$ is
\begin{equation}
    {\nabla _\pv} \overline \tau_i = \frac{1}{c}\left\| {{\pv_{\text{U}}} - {\pv_{\text{O}}}} \right\|^{-1} ({\pv_{\text{U}}} - {\pv_{\text{O}}})^\top,
\label{3-12}
\end{equation}
where $\text{O} \in \{\text{B,R}\}$ here depends on the type of channel geometrical parameters. Then, ${\nabla _\xv} \overline \zv^\top$ can be expressed as
\begin{equation}
{\nabla _\xv}\overline \zv^\top  = {\left[ {\begin{array}{*{20}{c}}
{{\nabla _\pv} \overline \tau_1}&{}\\
 \vdots &{}\\
{{\nabla _\pv} \overline \tau_N}&{}\\
{}&{{\nabla _{\text{vec}(\Rm)}} \overline \tv_1}\\
{}& \vdots \\
{}&{{\nabla _{\text{vec}(\Rm)}} \overline \tv_N}
\end{array}} \right]_{\left( {N + 6N} \right) \times \left( {3 + 9} \right)}}.
\label{3-13}
\end{equation}

\section{The Jacobian Matrix}\label{B}
In this appendix, we give the elements of the Jacobian matrix for the transformed covariance matrix in \eqref{JacobianMatrix}. 

The relation between $\pv$ and $\rv$ is given by \eqref{state} and \eqref{3-1-4}.
To facilitate the following calculations, we define:
\begin{equation}
   A_i = \displaystyle \frac{\partial (\displaystyle \frac{\sin \lambda }{\lambda }) }{\partial \rv_i} = \frac{\rv_i (\cos \lambda \cdot \lambda - \sin \lambda )}{\lambda^3} 
\label{A-1}
\end{equation}
\begin{equation}
B_i = \frac{\partial ( \displaystyle \frac{1 - \cos \lambda }{\lambda } )}{\partial \rv_i} = \frac{\rv_i (\sin \lambda \cdot \lambda + \cos \lambda -1 )}{\lambda^3} 
\label{A-2}
\end{equation}
\begin{equation}
C_{i,j} = \displaystyle \frac{\partial {\bm{\zeta}}_i }{\partial \rv_j} = \left\{\begin{matrix} 
-\lambda^{-3} \cdot \rv_i^2 + \displaystyle \frac{1}{\lambda }, \text{for} \ \ i = j 
 \\
-\lambda^{-3} \cdot \rv_i \cdot \rv_j , \text{for} \ \ i \ne  j 
\end{matrix}\right.
\label{A-3}
\end{equation}
\begin{equation}
D = {\bm{{\rho}}}^\top {\bm{\zeta}}, E = 1-\displaystyle \frac{\sin \lambda}{\lambda}, F = \displaystyle \frac{1- \cos \lambda}{\lambda}
\label{A-4}
\end{equation}
where $\lambda$ is the norm of $\rv$, and $\bm{\zeta}$ is the unit vector indicating the direction of $\rv$. $(\cdot)_i$ means the $i$-th element of the vector.

Each element in ${\partial \pv} \backslash {\partial {\rv}}$ can be calculated according to the following equation:
\begin{equation}
\begin{aligned}
\frac{\partial \mathbf{p}_1}{\partial \mathbf{r}_j} & =\boldsymbol{\rho}_1 \mathbf{a}_j-\mathbf{a}_j \zeta_1 D+E \zeta_1 \sum_{i=1}^3 \boldsymbol{\rho}_i \mathbf{C}_{i, j}+\mathbf{C}_{j, 1} D E  \\
& +\mathbf{b}_{j}\left(\zeta_2 \boldsymbol{\rho}_3-\zeta_3 \boldsymbol{\rho}_2\right)+F\left(\boldsymbol{\rho}_3 \mathbf{C}_{j, 2}-\boldsymbol{\rho}_2 \mathbf{C}_{j, 3}\right)
\label{A-11}
\end{aligned}
\end{equation}
\begin{equation}
\begin{aligned}
\frac{\partial \mathbf{p}_2}{\partial \mathbf{r}_j} & =\boldsymbol{\rho}_2 \mathbf{a}_j-\mathbf{a}_j \zeta_2 D+E \zeta_2 \sum_{i=1}^3 \boldsymbol{\rho}_i \mathbf{C}_{i, j}+\mathbf{C}_{j, 2} D E \\
& +\mathbf{b}_j\left(\zeta_3 \boldsymbol{\rho}_1-\zeta_1 \boldsymbol{\rho}_3\right)+F\left(\boldsymbol{\rho}_1 \mathbf{C}_{j, 3}-\boldsymbol{\rho}_3 \mathbf{C}_{j, 1}\right)  
\label{A-12}
\end{aligned}
\end{equation}
\begin{equation}
\begin{aligned}
\frac{\partial \mathbf{p}_3}{\partial \mathbf{r}_j} & =\boldsymbol{\rho}_3 \mathbf{a}_j-\mathbf{a}_j \zeta_3 D+E \zeta_3 \sum_{i=1}^3 \boldsymbol{\rho}_i \mathbf{C}_{i, j}+\mathbf{C}_{j, 3} D E \\
& +\mathbf{b}_j\left(\zeta_1 \boldsymbol{\rho}_2-\zeta_2 \boldsymbol{\rho}_1\right)+F\left(\boldsymbol{\rho}_2 \mathbf{C}_{j, 1}-\boldsymbol{\rho}_1 \mathbf{C}_{j, 2}\right)
\label{A-13}
\end{aligned}
\end{equation}
\normalsize

\end{appendices}


\bibliographystyle{IEEEtran}
\bibliography{IEEEabrv, ref}

\begin{thebibliography}{10}
\providecommand{\url}[1]{#1}
\csname url@samestyle\endcsname
\providecommand{\newblock}{\relax}
\providecommand{\bibinfo}[2]{#2}
\providecommand{\BIBentrySTDinterwordspacing}{\spaceskip=0pt\relax}
\providecommand{\BIBentryALTinterwordstretchfactor}{4}
\providecommand{\BIBentryALTinterwordspacing}{\spaceskip=\fontdimen2\font plus
\BIBentryALTinterwordstretchfactor\fontdimen3\font minus
  \fontdimen4\font\relax}
\providecommand{\BIBforeignlanguage}[2]{{%
\expandafter\ifx\csname l@#1\endcsname\relax
\typeout{** WARNING: IEEEtran.bst: No hyphenation pattern has been}%
\typeout{** loaded for the language `#1'. Using the pattern for}%
\typeout{** the default language instead.}%
\else
\language=\csname l@#1\endcsname
\fi
#2}}
\providecommand{\BIBdecl}{\relax}
\BIBdecl

\bibitem{chen2022tutorial}
H.~Chen, H.~Sarieddeen, T.~Ballal, H.~Wymeersch, M.-S. Alouini, and T.~Y.
  Al-Naffouri, ``A tutorial on terahertz-band localization for {6G}
  communication systems,'' \emph{{IEEE} Commun. Surveys Tuts.}, vol.~24, no.~3,
  pp. 1780--1815, thirdquarter. 2022.

\bibitem{huang2023virtual}
X.~Huang, J.~Riddell, and R.~Xiao, ``Virtual reality telepresence: 360-degree
  video streaming with edge-compute assisted static foveated compression,''
  \emph{{IEEE} Trans. Vis. Comput. Graphics}, vol.~29, no.~11, pp. 4525--4534,
  Nov. 2023.

\bibitem{alkhateeb2023real}
A.~Alkhateeb, S.~Jiang, and G.~Charan, ``Real-time digital twins: {Vision} and
  research directions for {6G} and beyond,'' \emph{IEEE Commun. Mag.}, vol.~61,
  no.~11, pp. 128--134, Nov. 2023.

\bibitem{Shahmansoori2017position}
A.~Shahmansoori, G.~E. Garcia, G.~Destino, G.~Seco-Granados, and H.~Wymeersch,
  ``Position and orientation estimation through millimeter-wave {MIMO} in {5G}
  systems,'' \emph{IEEE Trans. Wireless Commun.}, vol.~17, no.~3, pp.
  1822--1835, 2017.

\bibitem{Guerra2018Single}
A.~Guerra, F.~Guidi, and D.~Dardari, ``Single-anchor localization and
  orientation performance limits using massive arrays: {MIMO} vs.
  beamforming,'' \emph{IEEE Trans. Wireless Commun.}, vol.~17, no.~8, pp.
  5241--5255, 2018.

\bibitem{Abu2018Error}
Z.~Abu-Shaban, X.~Zhou, T.~Abhayapala, G.~Seco-Granados, and H.~Wymeersch,
  ``Error bounds for uplink and downlink {3D} localization in {5G} millimeter
  wave systems,'' \emph{IEEE Trans. Wireless Commun.}, vol.~17, no.~8, pp.
  4939--4954, 2018.

\bibitem{nazari2023mmwave}
M.~A. Nazari, G.~Seco-Granados, P.~Johannisson, and H.~Wymeersch, ``Mmwave {6D}
  radio localization with a snapshot observation from a single {BS},''
  \emph{{IEEE} Trans. Veh. Technol.}, vol.~72, no.~7, pp. 8914--8928, Jul.
  2023.

\bibitem{salem2025indoor}
H.~Salem, M.~Ahmed, M.~AlSharif, A.~Muqaibel, and T.~Al-Naffouri, ``Indoor
  position and attitude tracking with so (3) manifold,'' \emph{arXiv preprint
  arXiv:2501.01555}, 2025.

\bibitem{zheng2023coverage}
P.~Zheng, T.~Ballal, H.~Chen, H.~Wymeersch, and T.~Y. Al-Naffouri, ``Coverage
  analysis of joint localization and communication in {THz} systems with {3D}
  arrays,'' \emph{IEEE Trans. Wireless Commun.}, vol.~23, no.~5, pp.
  5232--5247, May. 2024.

\bibitem{Talvitie2023orientation}
J.~Talvitie, M.~Säily, and M.~Valkama, ``Orientation and location tracking of
  {XR} devices: {5G} carrier phase-based methods,'' \emph{IEEE J. Sel. Topics
  Signal Process.}, vol.~17, no.~5, pp. 919--934, Sept. 2023.

\bibitem{huang2020global}
Y.~Huang and Z.~Meng, ``Global distributed attitude tracking control of
  multiple rigid bodies via quaternion-based hybrid feedback,'' \emph{IEEE
  Trans. Control Netw. Syst.}, vol.~8, no.~1, pp. 367--378, Mar. 2020.

\bibitem{suvorova2021tracking}
S.~Suvorova, S.~D. Howard, and B.~Moran, ``Tracking rotations using maximum
  entropy distributions,'' \emph{IEEE Trans. Aerosp. Electron. Syst.}, vol.~57,
  no.~5, pp. 2953--2968, Oct. 2021.

\bibitem{10103759}
J.~Tian, M.~Yang, X.~Li, S.~Zou, and T.~Chao, ``Improvement of {RSS}-based
  measurement based on adaptive {Kalman} filter considering the anisotropy on
  antenna in dynamic environment,'' \emph{IEEE Trans. Instrum. Meas.}, vol.~72,
  pp. 1--13, 2023.

\bibitem{9893405}
B.~Yang, E.~Yang, L.~Yu, and C.~Niu, ``Adaptive extended {Kalman} filter-based
  fusion approach for high-precision {UAV} positioning in extremely confined
  environments,'' \emph{IEEE/ASME Trans. Mechatron.}, vol.~28, no.~1, pp.
  543--554, 2023.

\bibitem{varadarajan2013lie}
V.~S. Varadarajan, \emph{{Lie} groups, {Lie} algebras, and their
  representations}.\hskip 1em plus 0.5em minus 0.4em\relax Springer Science \&
  Business Media, 2013.

\bibitem{wang2006error}
Y.~Wang and G.~S. Chirikjian, ``Error propagation on the {Euclidean} group with
  applications to manipulator kinematics,'' \emph{IEEE Trans. Robot.}, vol.~22,
  no.~4, pp. 591--602, Aug. 2006.

\bibitem{bourmaud2015continuous}
G.~Bourmaud, R.~M{\'e}gret, M.~Arnaudon, and A.~Giremus, ``Continuous-discrete
  extended {Kalman} filter on matrix {Lie} groups using concentrated {Gaussian}
  distributions,'' \emph{J. Math. Imaging Vis.}, vol.~51, pp. 209--228, Jan.
  2015.

\bibitem{barfoot2014associating}
T.~D. Barfoot and P.~T. Furgale, ``Associating uncertainty with
  three-dimensional poses for use in estimation problems,'' \emph{IEEE Trans.
  Robot.}, vol.~30, no.~3, pp. 679--693, Jun. 2014.

\bibitem{barrau2014intrinsic}
A.~Barrau and S.~Bonnabel, ``Intrinsic filtering on {Lie} groups with
  applications to attitude estimation,'' \emph{IEEE Trans. Autom. Control},
  vol.~60, no.~2, pp. 436--449, Feb. 2014.

\bibitem{tao2022adaptive}
Z.~Tao, J.~Xue, D.~Wang, G.~Li, and J.~Fang, ``An adaptive invariant {EKF} for
  map-aided localization using {3D} point cloud,'' \emph{IEEE Trans. Intell.
  Transp. Syst.}, vol.~23, no.~12, pp. 24\,057--24\,070, Dec. 2022.

\bibitem{xu2023distributed}
J.~Xu, P.~Zhu, Y.~Zhou, and W.~Ren, ``Distributed invariant extended {Kalman}
  filter using {Lie} groups: Algorithm and experiments,'' \emph{IEEE Trans.
  Control Syst. Technol.}, Nov. 2023.

\bibitem{zhang2022adaptive}
H.~Zhang, D.~Ye, Y.~Xiao, and Z.~Sun, ``Adaptive control on {SE(3)} for
  spacecraft pose tracking with harmonic disturbance and input saturation,''
  \emph{IEEE Trans. Aerosp. Electron. Syst.}, vol.~58, no.~5, pp. 4578--4594,
  Oct. 2022.

\bibitem{petersen2022tracking}
M.~E. Petersen, J.~C. Ellingson, and R.~W. Beard, ``Tracking multiple unmanned
  aerial vehicles on {SE(3)} using a monocular camera,'' \emph{IEEE Trans.
  Aerosp. Electron. Syst.}, Aug. 2022.

\bibitem{stoica1998on}
P.~Stoica and B.~C. Ng, ``On the {Cram{\'e}r-Rao} bound under parametric
  constraints,'' \emph{IEEE Signal Process. Lett.}, vol.~5, no.~7, pp.
  177--179, Jul. 1998.

\bibitem{smith2005covariance}
S.~T. Smith, ``Covariance, subspace, and intrinsic {Cram{\'e}r-Rao} bounds,''
  \emph{IEEE Trans. Signal Process.}, vol.~53, no.~5, pp. 1610--1630, May.
  2005.

\bibitem{Nazari20213d}
M.~A. Nazari, G.~Seco-Granados, P.~Johannisson, and H.~Wymeersch, ``{3D}
  orientation estimation with multiple {5G} mmwave base stations,'' in
  \emph{Proc. IEEE Int. Conf. Commun. (ICC)}.\hskip 1em plus 0.5em minus
  0.4em\relax IEEE, 2021, pp. 1--6.

\bibitem{Boumal2013on}
N.~Boumal, ``On intrinsic {Cram{\'e}r-Rao} bounds for {Riemannian} submanifolds
  and quotient manifolds,'' \emph{IEEE Trans. Signal Process.}, vol.~61, no.~7,
  pp. 1809--1821, Apr. 2013.

\bibitem{labsir2024cramer}
S.~Labsir, S.~El~Bouch, A.~Renaux, J.~Vil{\`a}-Valls, and E.~Chaumette,
  ``Cram{\'e}r-rao bound for {Lie} group parameter estimation with euclidean
  observations and unknown covariance matrix,'' \emph{{IEEE} Trans. Signal
  Process.}, 2024.

\bibitem{wymeersch2022radio}
H.~Wymeersch and G.~Seco-Granados, ``Radio localization and sensing—part {I}:
  {Fundamentals},'' \emph{IEEE Commun. Lett.}, vol.~26, no.~12, pp. 2816--2820,
  Dec. 2022.

\bibitem{keskin2023esprit}
M.~F. Keskin, A.~Fascista, F.~Jiang, A.~Coluccia, G.~Seco-Granados, and
  H.~Wymeersch, ``{ESPRIT}-oriented precoder design for mmwave channel
  estimation,'' in \emph{Proc. IEEE Int. Conf. Commun. (ICC)}.\hskip 1em plus
  0.5em minus 0.4em\relax IEEE, 2023, pp. 903--908.

\bibitem{9965430}
T.~Ma, Y.~Xiao, and X.~Lei, ``Channel reconstruction-aided {MUSIC} algorithms
  for joint {AoA\&AoD} estimation in {MIMO} systems,'' \emph{IEEE Wirel.
  Commun. Lett.}, vol.~12, no.~2, pp. 322--326, 2023.

\bibitem{10273424}
X.~Zhang, H.~Zhang, and Y.~C. Eldar, ``Near-field sparse channel representation
  and estimation in {6G} wireless communications,'' \emph{IEEE Trans. Commun.},
  vol.~72, no.~1, pp. 450--464, 2024.

\bibitem{humphreys2012introduction}
J.~E. Humphreys, \emph{Introduction to {Lie} algebras and representation
  theory}.\hskip 1em plus 0.5em minus 0.4em\relax Springer Science \& Business
  Media, 2012, vol.~9.

\bibitem{legnani1996homogeneous}
G.~Legnani, F.~Casolo, P.~Righettini, and B.~Zappa, ``A homogeneous matrix
  approach to {3D} kinematics and dynamics—{I}. theory,'' \emph{Mech. Mach.
  Theory}, vol.~31, no.~5, pp. 573--587, 1996.

\bibitem{barfoot2024state}
T.~D. Barfoot, \emph{State estimation for robotics}.\hskip 1em plus 0.5em minus
  0.4em\relax Cambridge University Press, 2024.

\bibitem{dai2015euler}
J.~S. Dai, ``Euler--rodrigues formula variations, quaternion conjugation and
  intrinsic connections,'' \emph{Mech. Mach. Theory}, vol.~92, pp. 144--152,
  2015.

\bibitem{cardoso2010exponentials}
J.~R. Cardoso and F.~S. Leite, ``Exponentials of skew-symmetric matrices and
  logarithms of orthogonal matrices,'' \emph{J. Comput. Appl. Math.}, vol. 233,
  no.~11, pp. 2867--2875, 2010.

\bibitem{shen2020simultaneous}
S.~Shen, S.~Li, and H.~Steendam, ``Simultaneous position and orientation
  estimation for visible light systems with multiple {LEDs} and multiple
  {PDs},'' \emph{{IEEE} J. Sel. Areas Commun.}, vol.~38, no.~8, pp. 1866--1879,
  Aug. 2020.

\bibitem{kay1993fundamentals}
S.~M. Kay, \emph{Fundamentals of statistical signal processing: estimation
  theory}.\hskip 1em plus 0.5em minus 0.4em\relax Prentice-Hall, Inc., 1993.

\bibitem{boumal2023introduction}
N.~Boumal, \emph{An introduction to optimization on smooth manifolds}.\hskip
  1em plus 0.5em minus 0.4em\relax Cambridge University Press, 2023.

\bibitem{1668246}
Y.~Wang and G.~Chirikjian, ``Error propagation on the euclidean group with
  applications to manipulator kinematics,'' \emph{IEEE Trans. Robot.}, vol.~22,
  no.~4, pp. 591--602, 2006.

\bibitem{wolfe2011bayesian}
K.~C. Wolfe, M.~Mashner, and G.~S. Chirikjian, ``Bayesian fusion on {Lie}
  groups,'' \emph{Journal of Algebraic Statistics}, vol.~2, no.~1, 2011.

\bibitem{mahalanobis2018generalized}
P.~C. Mahalanobis, ``On the generalized distance in statistics,''
  \emph{Sankhy{\=a}: The Indian Journal of Statistics, Series A (2008-)},
  vol.~80, pp. S1--S7, 2018.

\bibitem{madyastha2011extended}
V.~Madyastha, V.~Ravindra, S.~Mallikarjunan, and A.~Goyal, ``Extended {Kalman}
  filter vs. error state {Kalman} filter for aircraft attitude estimation,'' in
  \emph{AIAA Guidance, Navigation, and Control Conference}, 2011, p. 6615.

\bibitem{ribeiro2004kalman}
M.~I. Ribeiro, ``Kalman and extended {Kalman} filters: Concept, derivation and
  properties,'' \emph{Institute for Systems and Robotics}, vol.~43, no.~46, pp.
  3736--3741, 2004.

\end{thebibliography}

\end{document}